\documentclass[
reprint,
superscriptaddress,
showpacs,
nofootinbib,
amsmath,amssymb,
aps,
prc,
floatfix]
{revtex4-1}

\usepackage{graphicx}
\usepackage{bm}

\usepackage{braket}

\begin{document}

\title{
Self-consistent band calculation of slab phase
in neutron-star crust
}

\author{Yu Kashiwaba}%
\email{kashiwaba@nucl.ph.tsukuba.ac.jp}
\affiliation{Faculty of Pure and Applied Sciences,
              University of Tsukuba, Tsukuba 305-8577, Japan}

\author{Takashi Nakatsukasa}%
\email{nakatsukasa@nucl.ph.tsukuba.ac.jp}
\affiliation{Center for Computational Sciences,
              University of Tsukuba, Tsukuba 305-8577, Japan}
\affiliation{Faculty of Pure and Applied Sciences,
              University of Tsukuba, Tsukuba 305-8577, Japan}
\affiliation{RIKEN Nishina Center, Wako 351-0198, Japan}

\date{\today}

\begin{abstract}
Fully self-consistent band calculation has been performed
for slab phase in neutron-star inner crust,
using the BCPM energy density functional.
Optimized slab structure is calculated at given baryon density
either with the fixed proton ratio or with the beta-equilibrium condition.
Numerical results indicate the band gap of in order of keV to tens of keV,
and the mobility of dripped neutrons are enhanced by the Bragg scattering,
which leads to the macroscopic effective mass,
$\bar{m}^*_z/m_n=0.65\sim 0.75$
near the bottom of the inner crust in neutron stars.
We also compare the results of the band calculation with those of
the Thomas-Fermi approximation.
The Thomas-Fermi approximation becomes invalid
at low density with high proton ratio.
\end{abstract}


\maketitle

\section{\label{sec:intro}Introduction}

Compact stars, such as white dwarfs and neutron stars, provide
extreme phases of matters.
In cold white dwarfs, the materials are ionized by high pressure,
producing the degenerate Fermi gas of electrons and crystalized ions.
In vicinity of the surface of neutron stars, called outer crust,
similar structure are expected to exist.
A little deeper in the neutron stars with greater density,
namely at inner crust, neutrons also form a degenerate Fermi gas.
We expect that the neutrons are in the $^1S_0$ pair condensation phase,
showing superfluidity.
There, neutron-rich nuclei form a Coulomb crystal which
coexist with gas of superfluid neutrons and electrons.
Even deeper in the stars, before the neutron-rich nuclei are
dissolved to form uniform nuclear matter (``core''),
we expect exotic phases of nuclear matter, called nuclear pasta
\cite{RPW83,HSY84,Oya93}.

In the pasta phase, the competition between the repulsive Coulomb
and the attractive nuclear interactions rearrange the nuclei
into exotic non-spherical shapes.
These shapes are often called name of pasta, such as spaghetti
(rod shape) and lasagna (slab shape).
These phases of non-spherical nuclei were first predicted using
a simple liquid drop model \cite{RPW83,HSY84}.
The Thomas-Fermi approximation to a simplified Skyrme-like energy
density functional confirmed the stability of these phases \cite{Oya93}.
Since the pasta phase has a regular structure analogous to the crystal,
quantum mechanically, it should be treated as a system with
a periodic potential, namely, with the Bloch boundary condition.
Chamel has performed the band calculation to predict
large effective mass $m^*$ of neutrons due to the Bragg scattering by
the periodic potential produced by the crystalline structure \cite{Cha12}.
The calculation predicts that
the ratio $m^*/m_n$, where $m_n$ is the bare neutron mass,
can be 10 or even larger.

The structure of the inner crust has a significant influence on
observed properties of neutron stars.
Especially, the entrainment effect could have a significant
consequence on an interpretation of glitch phenomena.
The pulsar glitches first were observed in 1969 at the Vela pulsar
\cite{RM69,RD69},
in which the spin rate of the pulsar suddenly increases.
Since then, we have observed more than 100 pulsars showing
the glitch phenomena.
The current consensus view is that
the vortices in the superfluid neutron gas in the inner crust
are responsible for the glitches \cite{AI75}.
The spin-rate difference between the superfluid neutrons and
the rest of the star is accumulated until the glitch occurs.
If the superfluid neutrons in the inner crust
are a reservoir of the angular momentum,
in order to explain the magnitude of glitches,
the ratio between the neutrons' and the total moments of inertia
should be $I_n/I\sim 1-1.5$ \% \cite{LEL99}.
These values are consistent with 
typical nuclear matter equations of state (EOS).
However, taking into account the entrainment effect,
the required ratio $I_n/I$ is multiplied by $m^*/m_n$,
which leads to serious difficulties to regard 
the superfluid neutrons as the angular momentum reservoir
\cite{AGHE12}.

The structure of the inner crust also
influences the electrical and the thermal conductivity.
A high electrical resistive layer in the bottom part of the inner crust,
supposed to be the pasta layers,
could lead to magnetic field decay of neutron stars.
This limits the maximum spin period,
which is consistent with lack of observation of isolated X-ray pulsars
with spin period longer than 12 s \cite{PVR13}.
Since the thermal conductivity is likely related to the electrical one,
the effect of the nuclear pasta may be also seen in the late time crust
cooling \cite{Hor15}.

The calculation by Chamel \cite{Cha12} predicted that the density of 
conduction neutrons $n^c_n$ is significantly smaller than the density of
unbound neutrons $n^f_n$, leading to the large effective mass,
$m^*/m_n = n_n^f/n_n^c$.
This may require us to revise our interpretation of pulsar glitches.
However, the calculation of Ref.~\cite{Cha12} is not self-consistent.
He adopted a periodic one-body potential based on the extended Thomas-Fermi
calculation in the Wigner-Seitz cell.
In this paper, we perform the band calculation using a modern energy density functional
in a fully self-consistent manner.
As the first step toward the systematic calculation of various phases,
we treat the slab phase, neglecting spin-orbit and pairing correlations.

The paper is organized as follows.
In Sec.~\ref{sec:band-theory}, we give the formulation of the
self-consistent calculation of slab nuclei with dripped neutrons
using the band theory.
We perform numerical calculations for systems
with a given proton ratio and at the beta-equilibrium condition.
The entrainment effect is measured by effective mass.
In Sec.~\ref{sec:numerical},
we show results of the numerical calculations.
For comparison, we also perform calculation
with the Thomas-Fermi approximation which has been frequently used
in the past.
Summary and future perspectives are given in Sec.~\ref{sec:summary}.

\section{Band theory for slab phase}
\label{sec:band-theory}

The inner crust of neutron stars have been studied quantum mechanically
with the Wigner-Seitz approximation, in which
the periodic structure is decomposed into independent spherical cells
of a cell radius $R_{\rm cell}$.
Unbound neutrons are treated with different boundary conditions
at $R_{\rm cell}$ according to the parity of the orbitals \cite{NV73}.
The Wigner-Seitz approximation is relatively well justified
for the outer crust, while it is questionable for the inner crust \cite{BST06}.
Especially, the entrainment effect cannot be taken into account in
this approximation.

\subsection{Reduction to one dimension}

\begin{figure}[tb]
    \centerline{
\includegraphics[width=\columnwidth,pagebox=cropbox,clip]{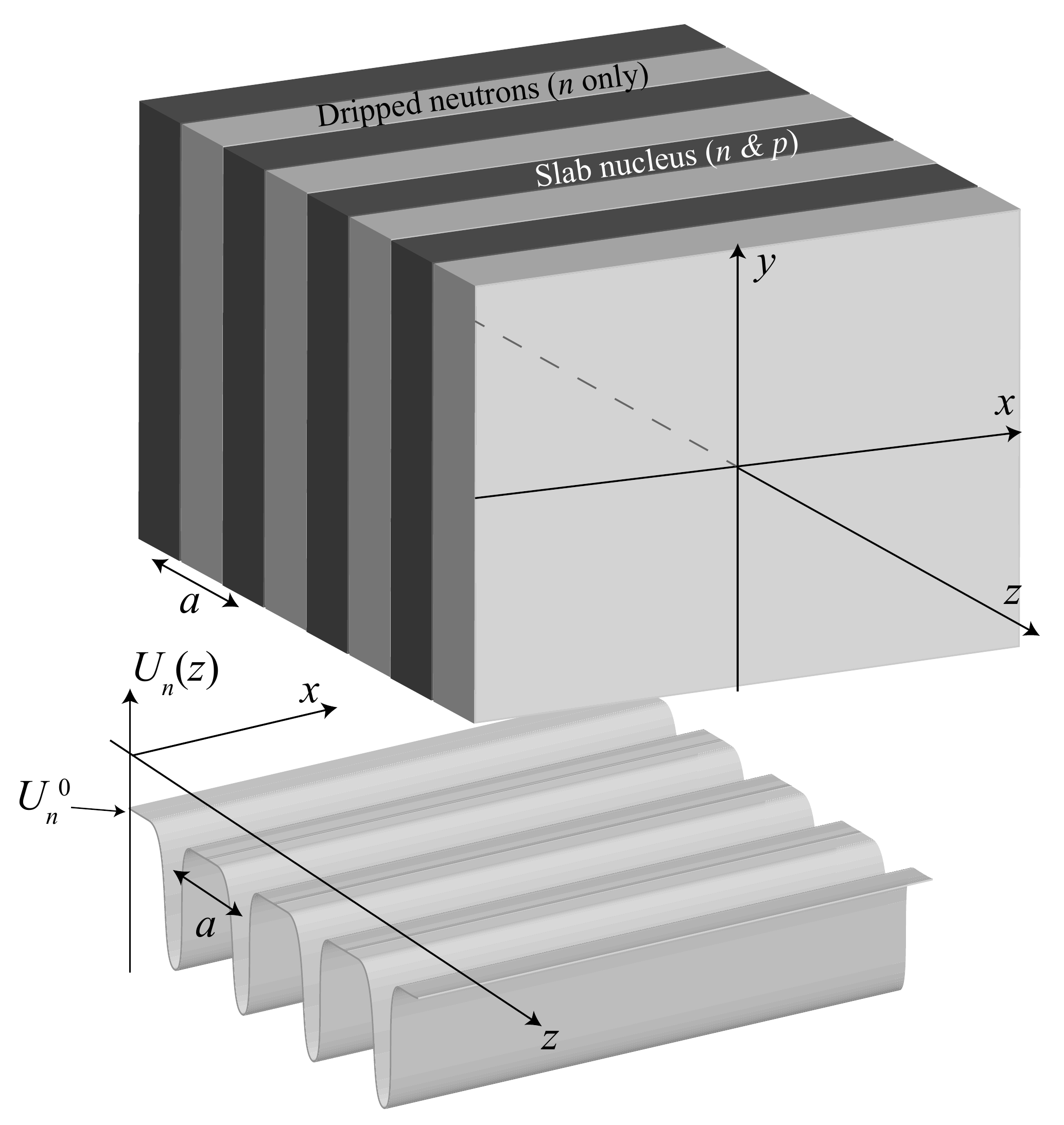}
}
\caption{
(Top) Schematic picture of the slab phase.
The slabs are parallel to the $x$-$y$ plane.
(Bottom) 
Schematic illustration of potential for neutrons $U_n(z)$
in the $z$-$x$ plane.
}
\label{fig:slab_phase}
\end{figure}

The proper treatment of unbound neutrons in a crystallized nuclear matter
requires the band theory of solids \cite{AM76}.
Although the band theory is a well-known established theory,
as far as we know,
its application to neutron-star crust was first performed by Chamel in 2005
\cite{Cha05}.
The essential difference from the treatment of an isolated nucleus
is the boundary condition for the single-particle (Kohn-Sham)
wave functions.
They satisfy the Bloch boundary condition \cite{AM76}:
\begin{equation}
\varphi_{\alpha,\bm{k}}^{(q)}(\bm{r}+\bm{T},s)
= \exp(i\bm{k}\cdot\bm{T})
\varphi_{\alpha,\bm{k}}^{(q)}(\bm{r},s) ,
\label{Bloch_theorem}
\end{equation}
where $\bm{k}$ is the Bloch wave vector and $\alpha$ is the band index.
The vector $\bm{T}$ is a lattice translational vector.
The spin and isospin indices are denoted by
$s=\pm 1/2$ and $q=(n,p)$.
We focus the present work on the slab phase,
and choose
the normal direction to the slab nuclei as the $z$ direction.
See Fig.~\ref{fig:slab_phase}.
The system is assumed to be uniform along the tangential 
($x$-$y$) directions,
and the spin-orbit potential is neglected for simplicity.
Assuming the distance $a$ between neighboring slab nuclei,
the lattice vector becomes $\bm{T}=T_x\bm{e}_x+T_y\bm{e}_y+na\bm{e}_z$,
where $T_x$ and $T_y$ are arbitrary real numbers,
$n$ is an arbitrary integer number, and
$\bm{e}_j$ is a unit vector along $j=(x,y,z)$ direction.
The wave functions become separable, in which
those with respect to $x$ and $y$ coordinates are the plane waves.
Thus, we may write the wave function
$\varphi_{\alpha,\bm{k}}^{(q)}(\bm{r})$
of Eq. (\ref{Bloch_theorem}) in a form
\begin{equation}
\label{separable_wf}
\varphi_{\alpha,\bm{k}}^{(q)}(\bm{r})=
\exp(ik_x x+ik_y y) \cdot e^{ik_z z}
\phi_{\alpha,\bm{k}}^{(q)}(z)
=e^{i\bm{k}\cdot\bm{r}}
\phi_{\alpha,\bm{k}}^{(q)}(z) ,
\end{equation}
where the wave functions $\phi_{\alpha,\bm{k}}^{(q)}(z)$
satisfy the periodic boundary condition
\begin{equation}
\phi_{\alpha,\bm{k}}^{(q)}(z+a) =
\phi_{\alpha,\bm{k}}^{(q)}(z) .
\end{equation}
Hereafter, we omit the index $s$, because the wave functions are
spin-independent.
Thus, the model space can be reduced into
the one-dimensional unit cell, $0\leq z < a$.
Here, $\bm{k}=k_x\bm{e}_x + k_y\bm{e}_y + k_z\bm{e}_z$, and
$(k_x,k_y)$ represents the nucleon momentum in the tangential
($x$-$y$) directions,
while $k_z$ corresponds to the Bloch wave number
which can be restricted inside the first Brillouin zone,
$-\pi/a < k_z \leq \pi/a$.

The wave functions are determined by the self-consistent (Kohn-Sham) equations,
\begin{equation}
h_q[\rho] \varphi_{\alpha,\bm{k}}^{(q)}(\bm{r})
= \epsilon_{\alpha,\bm{k}}^{(q)} \varphi_{\alpha,\bm{k}}^{(q)}(\bm{r}) ,
\end{equation}
with the single-particle Hamiltonian
\begin{equation}
h_q[\rho] = -\nabla \cdot \frac{1}{2m_q^*(\bm{r})} \nabla + U_q(\bm{r}) .
\end{equation}
In this paper, we adopt the unit of $\hbar=1$.
In the present case, since the slab is assumed to be uniform in the
the effective mass $m_q^*(\bm{r})$ and 
the selfconsistent potential $U_q(\bm{r})$ depends only on $z$.
It should be noted that the effective mass $m_q^*(\bm{r})$
here is different from those discussed in
Secs.~\ref{sec:mobility} and \ref{sec:numerical}.
When we need to distinguish different effective masses in the present paper,
we call $m_q^*(\bm{r})$ {\it microscopic} effective mass, and those
in Sec.~\ref{sec:mobility} {\it macroscopic} ones.
The wave functions of Eq. (\ref{separable_wf}) lead to
\begin{equation}
\left(\frac{k_\rho^2}{2m_q^*(z)}+h_{k_z}^{(q)} \right)
\phi_{\alpha,\bm{k}}^{(q)}(z) 
= \epsilon_{\alpha,\bm{k}}^{(q)} \phi_{\alpha,\bm{k}}^{(q)}(z) , 
\label{Bloch_eq_1}
\end{equation}
with $k_\rho^2\equiv k_x^2+k_y^2$ and
\begin{equation}
h_{k_z}^{(q)}\equiv 
\left(-i\partial_z + k_z \right) \frac{1}{2m_q^*(z)}
\left(-i\partial_z + k_z \right) +U_q(z)  .
\label{sp_Hamiltonian_1}
\end{equation}

We end up at the one-dimensional equation (\ref{Bloch_eq_1}),
though the wave functions still depend on $(k_x,k_y)$
through the $z$-dependent kinetic energy term.
If the effective mass $m_q^*(z)$ is identical to the bare nucleon mass
$m_q$, the single-particle wave functions $\phi_{\alpha,\bm{k}}^{(q)}(z)$
become independent from $(k_x,k_y)$;
$\phi_{\alpha,\bm{k}}^{(q)} \rightarrow \phi_{\alpha,k_z}^{(q)}$.
Since this significantly reduces computational task,
in this study,
we adopt an energy density functional with no derivative terms
(Sec.~\ref{sec:EDF}).
Equation (\ref{Bloch_eq_1}) is reduced to
\begin{equation}
h_{k_z}^{(q)} \phi_{\alpha,k_z}^{(q)}(z) 
= e_{\alpha,k_z}^{(q)} \phi_{\alpha,k_z}^{(q)}(z) , 
\label{Bloch_eq_2}
\end{equation}
where $e_{\alpha,k_z}^{(q)}$ represent the energy band as functions of $k_z$.
The single-particle energy is given by
\begin{equation}
\epsilon_{\alpha,\bm{k}}^{(q)} = \frac{k_\rho^2}{2m_q} + e_{\alpha,k_z}^{(q)} .
\label{sp_energy}
\end{equation}
The states with $\epsilon_{\alpha,\bm{k}}^{(q)}<\mu_q$ are occupied,
where $\mu_q$ is the Fermi energy (chemical potential).

In practice, the Bloch wave number, $-\pi/a < k_z \leq \pi/a$
is discretized into $N_k$ points.
The calculation with $N_k$ points of $k_z$ is identical to
the calculation with the periodic boundary condition
in a space $N_k$ times larger than the unit cell.
The wave functions are normalized as
\begin{equation}
\int_0^a \left| \phi_{\alpha,k_z}^{(q)}(z)\right|^2 dz = \frac{1}{N_k} .
\end{equation}
The density is calculated as
\begin{eqnarray}
\rho_q(z)&=&2\sum_{\alpha,k_z} \left| \phi_{\alpha,k_z}^{(q)}(z)\right|^2
\int \frac{dk_x dk_y}{(2\pi)^2}
\theta(\mu_q-\epsilon_{\alpha,\bm{k}}^{(q)}) \nonumber \\
    &=&\frac{m_q}{\pi} \sum_{\alpha,k_z}^{\rm occ}
    \left| \phi_{\alpha,k_z}^{(q)}(z)\right|^2 
\left(\mu_q-e_{\alpha,k_z}^{(q)}\right) ,
    \label{rho_q}
\end{eqnarray}
where $\theta(x)$ is a step function and
$\sum^{\rm occ}$ means the summation with respect to
occupied (hole) orbitals only.
The summation with respect to $k_z$ is taken over $N_k$ discretized
values of $k_z$.
The baryon (nucleon) number density is defined as
$\rho_B(z)\equiv \rho_n(z)+\rho_p(z)$.
In a similar manner, the kinetic density is given by
\begin{eqnarray}
\tau_q(z) &\equiv&
2 \sum_{\alpha,k_z} \int \frac{dk_x dk_y}{(2\pi)^2}
\left| \nabla \varphi_{\alpha,\bm{k}}^{(q)}(\bm{r}) \right|^2
\theta(\mu_q-\epsilon_{\alpha,\bm{k}}^{(q)})
\nonumber \\
&=&  \frac{m_q^2}{\pi} \sum_{\alpha,k_z}^{\rm occ}
\left(\mu_q-e_{\alpha,k_z}^{(q)}\right)^2
\left| \phi_{\alpha,k_z}^{(q)}(z)\right|^2 \nonumber \\
&+&\frac{m_q}{\pi} \sum_{\alpha,k_z}^{\rm occ}
\left(\mu_q-e_{\alpha,k_z}^{(q)}\right) 
\left| \left(ik_z+\frac{d}{dz}\right)
\phi_{\alpha,k_z}^{(q)}\right|^2 .
    \label{tau_q}
\end{eqnarray}

\subsection{Nuclear and electronic energy density functionals}
\label{sec:EDF}

The average baryon number density $n_B$ is
given by $n_B=n_n+n_p$, with the average neutron and proton densities,
\begin{equation}
n_q \equiv \frac{1}{a} \int_0^a \rho_q(z)dz .
\end{equation}
The proton fraction $Y_p$ is defined by $Y_p\equiv n_p/n_B$.
In this study, the electrons are assumed to be uniform.
The charge neutrality requires that the electron density is
equal to the average proton density, $n_e=n_p$.
The charge density is simply given by
$\rho_c(\bm{r})\equiv \rho_p(\bm{r})-n_e$,
neglecting the charge form factor of protons.
The electrons are treated as degenerated relativistic Fermi gas
with the Fermi momentum, $p_F$.
The Fermi energy is given as
$\epsilon_F^{(e)}=\sqrt{m_e^2 c^4 + p_F^2 c^2} = m_e c^2 \cosh\theta_F$,
where $\theta_F$ is determined by
the electron density, $3 \pi^2 n_e=(m_e c \sinh\theta_F)^3$.
Their energy divided by the baryon (nucleon) number $A$
is given as $E_e/A=K_e/A+E_C^{(e)}/A$ with
\begin{eqnarray}
\label{K_e}
\frac{K_e}{A} &=& \frac{4\pi}{n_B} \int_0^{p_F} \frac{p^2dp}{(2\pi)^3}
    \sqrt{m_e^2c^4 + p^2c^2} \nonumber \\
    &=& \frac{m_e^4 c^5}{32\pi^2 n_B}
    \left( \sinh 4\theta_F - 4\theta_F \right) ,\\
\label{E_C^e}
\frac{E_C^{(e)}}{A} &=&
-\frac{3e^2 n_e^{4/3}}{4n_B}\left(\frac{3}{\pi}\right)^{1/3} ,
\end{eqnarray}
where we use the Slater approximation for the exchange energy.
The direct term of the Coulomb energy
$-(2an_B)^{-1}\int_0^a V_C(z) n_e dz$, vanishes,
because of the charge neutrality condition for
the Coulomb potential;
\begin{equation}
\int_0^a V_C(z)dz = 0 .
    \label{V_C_boundary_condition}
\end{equation}
The Coulomb potential for protons $V_C$ ($-V_C$ for electrons)
is calculated by solving the Poisson equation
\begin{equation}
\label{poisson}
\frac{d^2}{dz^2} V_C(z)=-4\pi e^2 \rho_c(z) ,
\end{equation}
with the condition (\ref{V_C_boundary_condition}).

For nuclear part,
we adopt the energy density functional (EDF) of BCPM \cite{BRSV13},
neglecting the spin-orbit coupling terms.
This EDF is constructed so as to reproduce properties of both
the nuclear matter, predicted by the Brueckner-Hartree-Fock calculation,
and experimental data of finite nuclei.
Another practical reason of this choice is that
the kinetic density $\tau_q(\bm{r})$
is present only in the kinetic energy terms.
Thus, we can use Eq. (\ref{Bloch_eq_2}) instead of
Eq. (\ref{Bloch_eq_1}), which significantly reduces
the computational cost.

In the BCPM functional, the nuclear energy is written as
\begin{equation}
E_B = \sum_{q=n,p} K_q + E_{\rm vol} + E_{\rm sur}
    + E_{\rm C}^{(p)} .
\label{E_B}
\end{equation}
Here, $K_q$ is the kinetic energy of neutrons ($q=n$) and protons ($q=p$).
The potential energy is divided into two parts,
the ``volume'' (bulk) and ``surface'' parts.
The volume part $E_{\rm vol}$ reproduces the nuclear matter
properties, and the surface part $E_{\rm sur}$ is added to
reproduce observed properties of finite nuclei \cite{BRSV13}.
The kinetic energy per baryon is given by
\begin{eqnarray}
    \label{K_q}
\frac{K_q}{A}=
\frac{1}{a n_B}\int_0^a \frac{\tau_n(z)+\tau_p(z)}{2m_q} dz .
\end{eqnarray}
The volume part is given as
\begin{eqnarray}
\frac{E_{\rm vol}}{A} &=&
    \frac{1}{an_B} \int_0^a \left[
P_s(\rho_B(z))\left\{1-\beta^2(z)\right\}
\right. \nonumber \\
    && \quad\quad\quad\quad
    \left. +P_n(\rho_B(z))\beta^2(z) \right] \rho_B(z) dz ,
\end{eqnarray}
where $\beta(z)\equiv (\rho_n(z)-\rho_p(z))/\rho_B(z)$.
The functionals $P_s(\rho_B)$ and $P_n(\rho_B)$ are given in
polynomial with respect to $\rho_B(z)$ \cite{BRSV13}.
The surface part is
\begin{eqnarray}
\label{E_sur}
\frac{E_{\rm sur}}{A} &=&
\frac{\pi r_0^2}{2an_B} \sum_{q,q'} V_{qq'}
    \left[ \int_0^a dz \int_0^a dz' \rho_q(z) e^{-\frac{(z-z')^2}{r_0^2}} \rho_{q'}(z')
\right. \nonumber \\
    && \quad\quad\quad\quad\quad\quad
    \left. -g \int_0^a dz \rho_q(z) \rho_{q'}(z) \right] .
\end{eqnarray}
This expression ignores the interaction with nucleons
outside of the unit cell.
In order to take into account interactions with
nucleons in the neighboring cells,
we replace the gaussian 
$e^{-(z-z')^2/r_0^2}$ in Eq. (\ref{E_sur}) by
\begin{equation}
v(z)\equiv
e^{-(z-z')^2/r_0^2}+e^{-(z-z'-a)^2/r_0^2}+e^{-(z-z'+a)^2/r_0^2} .
\end{equation}
This treatment is well justified at $a \gg r_0$. 
The parameter $g$ is given, in this approximation, by
$
g\equiv \int_0^a v(z) dz
$.
See Ref.~\cite{BRSV13} for values of the parameters $V_{qq'}$ and $r_0$.

The Coulomb energy per baryon is given by
\begin{eqnarray}
\label{E_C^p}
\frac{E_C^{(p)}}{A} &=& \frac{1}{2a n_B} \int_0^a V_C(z) \rho_p(z) dz
\nonumber \\
&& \quad
    -\frac{3e^2}{4an_B}\left(\frac{3}{\pi}\right)^{1/3}
\int_0^a \rho_p^{4/3}(z) dz .
\end{eqnarray}
The direct part of the total Coulomb energy is
\begin{equation}
	\left(E_C^{(e)}+E_C^{(p)}\right)_D =
\frac{e^2}{2} \iiint d^3\bm{r} \iiint d^3\bm{r}'
\frac{\rho_c(z)\rho_c(z')}{\left|\bm{r}-\bm{r}'\right|} .
\end{equation}

\subsection{Self-consistent solutions}
\label{sec:self-consistent_solutions}

The potentials $U_q(z)$
is given by the variation of the interaction energy with respect to
the density.
\begin{equation}
U_q(z)=\frac{\delta (E_{\rm vol}+E_{\rm sur}+E^{(q)}_C)}{\delta\rho_q(z)} ,
\end{equation}
where $E^{(n)}_C=0$.
Apparently, these potentials are functionals of $\rho_q(z)$,
and the self-consistency is required for solutions of Eq. (\ref{Bloch_eq_2}).
In the present case, since we have the effective mass of $m_q^*(z)=m_q$
the single-particle Hamiltonian of Eq. (\ref{sp_Hamiltonian_1}) is
\begin{equation}
h_{k_z}^{(q)}=
\frac{1}{2m_q}\left(-i\partial_z + k_z \right)^2 + U_q(z)  .
\label{sp_Hamiltonian_2}
\end{equation}

In the band calculation of solid, the Hamiltonian of electrons
explicitly depends on
the coordinates of ions ($\bm{R}$) and the structure optimization 
is performed using the Feynman-Hellman theorem,
$d\langle H \rangle/d\bm{R}
=\langle dH/d\bm{R} \rangle = 0$.
However, for the band calculation of neutron-star crust,
the Hamiltonian is independent from nuclear configuration.
The crystalline structure is defined ``by hand'' in the beginning,
in order to reduce the
large-space calculation into that of a unit cell.
Thus, the use of the Feynman-Hellman theorem is not trivial.
In the present paper, we perform the optimization
with respect to the lattice constant (slab interval) $a$,
by explicitly calculating different values of $a$.

The total energy of the system is simply given by the sum of
nuclear and electronic energies, $E_B + E_e$.
We perform calculations with a given value of
the average proton density $n_p$.
Thus, the uniform electron density $n_e=n_p$ is also fixed.
In order to find the optimum value of the lattice constant $a$,
we should minimize the total energy,
\begin{equation}
\left.\frac{\partial(E_B+E_e)}{\partial a}\right|_{n_p}
    = 
\left.\frac{\partial E_B}{\partial a}\right|_{n_p}
   = 0 .
\end{equation}
Therefore, the minimization of the nuclear energy gives
the optimal value of $a$.

\subsubsection{Fixed proton ratio}
\label{sec:fixed_n_p}

In an event of supernova, during collapse of a giant star,
the nuclear matter at
a large variety of density and proton ratio is supposed to appear.
When we fix the average neutron and proton densities,
the electron energy, $E_e=K_e+E_C^{(e)}$ of
Eqs. (\ref{K_e}) and (\ref{E_C^e}),
plays any role neither for solutions of
the self-consistency equation (\ref{Bloch_eq_2}),
nor for the optimization of the slab distance $a$.
Thus, we solve Eq. (\ref{Bloch_eq_2}) with $n_e=n_p$
for a given value of the lattice constant $a$.

The iterative calculation at a given $a$ is performed
according to the following procedure.
\begin{enumerate}
\item Input the initial density, $\rho_q^{(i)}(z)$ ($i=0$),
    and calculate the energy per baryon $E_B^{(i)}/A$.
\item \label{enu:pot_cal} Calculate the potential $U_q(z)$.
\item Solve Eq. (\ref{Bloch_eq_2}) to determine
    the wave functions $\phi_{\alpha,k_z}^{(q)}(z)$ and
    the band energy $e_{\alpha,k_z}^{(q)}$.
\item \label{enu:chem_pot}
    Determine the chemical potential $\mu_q$ so as to obtain
    the target density $n_q$.
\item Calculate the new densities,
$\rho_q(z)$ and $\tau_q(z)$,
according to Eqs. (\ref{rho_q}) and (\ref{tau_q}), respectively.
\item Calculate the nuclear energy $E_B/A$.
    Check the convergence condition, $|(E_B - E_B^{(i)})/E_B| < \eta$.
If this is not satisfied, update the density;
$\rho_q^{(i+1)}(z)=(1-\omega)\rho_q^{(i)}(z) + \omega \rho_q(z)$,
set $E_B^{(i+1)}=E_B$, and
go back to Step \ref{enu:pot_cal}.
\end{enumerate}
Here, we adopt the parameters, $\eta=10^{-10}$ and $\omega=0.1$.

\subsubsection{Beta equilibrium}

The electron energy is completely irrelevant for the self-consistent
solutions in the case that both $n_p$ and $n_n$ are fixed.
In contrast, it affects the condition of beta equilibrium.
The beta-equilibrium condition is given by
\begin{equation}
(\mu_p + m_p c^2) + \mu_e = (\mu_n + m_n c^2) .
\label{beta_equilibrium}
\end{equation}
Note that the electron chemical potential contains its rest mass,
\begin{equation}
\mu_e=\frac{1}{V}\frac{\partial E_e}{\partial n_e}
    =\sqrt{m_e^2 c^4 + p_F^2 c^2}
    -e^2 \left(\frac{3n_e}{\pi}\right)^{1/3} .
\end{equation}
We perform the calculation at a given average density of protons, $n_p$.
Since the charge neutral condition requires $n_e=n_p$,
the condition (\ref{beta_equilibrium}) determines the neutron
chemical potential.
Then, the iterative procedure (Steps $1\sim 6$) is exactly the same as
the previous one in Sec.~\ref{sec:fixed_n_p},
except for Step \ref{enu:chem_pot} where $\mu_n$ is given by the condition
$\mu_n=\mu_p+\mu_e-(m_n-m_p) c^2$.

\subsection{Entrainment and effective mass}
\label{sec:mobility}

In the outer crust, both neutrons and protons are bound in nuclei.
In the rest frame of crust, even with a perturbative force on
neutrons, there would be no current.
In contrast, the inner crust has conduction neutrons which
are dripped from the nuclear binding.
Thus, the band filling property determines whether they are
``conductor'' or ``insulator''.
We expect that the slab phase of the inner crust is always a conductor,
because the neutron single-particle energy
$\epsilon_{\alpha,\bm{k}}^{(n)}$ in  Eq. (\ref{sp_energy})
is continuous and has no gap.
Nevertheless, its $z$ component, $e_{\alpha,k_z}^{(n)}$,
represents the band structure and has band gaps at
$k_z=\pm \pi/a$, which may affect the conduction properties along
the $z$ direction (normal to the slabs).

The group velocity of neutrons in the band $\epsilon_{\alpha,\bm{k}}^{(n)}$
is given by \cite{AM76}
\begin{equation}
\bm{v}^{(n)}_{\alpha,\bm{k}}
=\nabla_{\bm{k}} \epsilon^{(n)}_{\alpha,\bm{k}} .
\label{group_velocity}
\end{equation}
Assuming no interband transition between bands with different $\alpha$,
the acceleration under a force $\bm{F}$ on neutrons is given by
\begin{equation}
	\frac{d\bm{v}^{(n)}_{\alpha,\bm{k}}}{dt}
=\left( \frac{d\bm{k}}{dt} \cdot\nabla_{\bm{k}} \right)
	\bm{v}_{\alpha,\bm{k}}^{(n)}
    =\left( \bm{F} \cdot\nabla_{\bm{k}} \right)
    \nabla_{\bm{k}} \epsilon^{(n)}_{\alpha,\bm{k}} ,
\end{equation}
where we used the acceleration theorem, $d\bm{k}/dt=\bm{F}$
\cite{Cal76,AM76}.
Writing the $i$-th component ($i=x,y,z$) of $\bm{v}$ as
\begin{equation}
\frac{dv^i}{dt}= \sum_j \left(\frac{1}{m^*} \right)_{\alpha,\bm{k}}^{ij} F_j ,
\end{equation}
we obtain the {\it macroscopic} effective mass tensor
\begin{equation}
    \left(\frac{1}{m^*} \right)^{ij}_{\alpha,\bm{k}}
\equiv
\frac{\partial^2 \epsilon^{(n)}_{\alpha,\bm{k}}}{\partial k_i\partial k_j} .
    \label{effective_mass_t}
\end{equation}
In the present case, the effective mass is diagonal
($(1/m^*)^{ij}=\delta^{ij}/m_i^{*}$),
and for $i=x,y$, they are equal to the bare neutron mass,
$m^{*}_i=m_n$.

The neutron mobility can be measured by
Eq. (\ref{effective_mass_t}) summed over the occupied orbits.
\begin{equation}
\mathcal{K}^{ij}\equiv
2 \sum_\alpha \int \frac{d^3k}{(2\pi)^3}
    \frac{\partial^2 \epsilon_{\alpha,\bm{k}}^{(n)}}
    {\partial k_i \partial k_j}
    \theta(\mu_n-\epsilon_{\alpha,\bm{k}}^{(n)} ) .
\label{mobility_coefficient}
\end{equation}
For the present case of the slab phase, it is transformed to
\begin{equation}
\mathcal{K}^{ij}\equiv
\frac{2}{a N_k} \sum_{\alpha,k_z} \int \frac{dk_x dk_y}{(2\pi)^2}
    \frac{\partial^2 \epsilon_{\alpha,\bm{k}}^{(n)}}
    {\partial k_i \partial k_j}
    \theta(\mu_n-\epsilon_{\alpha,\bm{k}}^{(n)} ) ,
\end{equation}
and $\mathcal{K}^{ij}$ is diagonal in
the Cartesian coordinates $(x,y,z)$, namely,
$\mathcal{K}^{ij}=0$ for $i\neq j$.
The mobility coefficients for $x$ and $y$ directions are
simply given as
$\mathcal{K}^{xx}=\mathcal{K}^{yy}=n_n/m_n$,
The $z$ component is calculated as
\begin{equation}
\mathcal{K}^{zz}\equiv
\frac{m_n}{\pi a N_k} \sum_{\alpha,k_z}^{\rm occ}
    \frac{d^2 e_{\alpha,k_z}^{(n)}}{dk_z^2}
    \left(\mu_n- e_{\alpha,k_z}^{(n)} \right) ,
\label{mobility_coefficient_z}
\end{equation}
and equivalently given by 
\begin{equation}
\mathcal{K}^{zz}\equiv
\frac{m_n}{\pi a N_k} \sum_{\alpha,k_z}^{\rm occ}
	\left(\frac{d e_{\alpha,k_z}^{(n)}}{dk_z}\right)^2
\label{mobility_coefficient_z_2}
\end{equation}
From this mobility coefficient, we may define 
conduction neutron density $n_i^c$
which are supposed to freely move in the neutron-star crust
along the $i$ direction ($i=x,y,z$)
\cite{Cha12}.
\begin{equation}
n_i^c \equiv m_n \mathcal{K}^{ii} .
\end{equation}
Trivially, we have $n_x^c = n_y^c = n_n$,
which means that all the neutrons in the slab phase
are effectively free in the $x$-$y$ plane.
In contrast, the $z$ component of the conduction neutron density $n_z^c$
may be hindered, not only by the bound neutrons inside the slab, but also
by the Bragg scattering due to the periodic nuclear potential.
The latter is called entrainment effect \cite{CCH05,CCH06,Cha12}.

The reduction of $n_z^c$ from the neutron density $n_n$ is quantified
as an effective mass \cite{CCH06}:
\begin{equation}
m_z^* \equiv \frac{n_n}{\mathcal{K}^{zz}} = m_n \frac{n_n}{n_z^c} ,
    \label{effective_mass_1}
\end{equation}
With this definition (\ref{effective_mass_1}), the effective mass
diverges in the outer crust where there are no dripped neutrons.

Another definition is given as the ratio to ``free'' neutrons.
In this paper, we adopt two kinds of definition.
The first one is given by
neutrons whose single-particle energy is larger than
the maximum value of the neutron potential $U_n^0\equiv \max (U_n(z))$
(See also Fig.~\ref{fig:slab_phase}).
With this definition, we have $n_n^f = 0$ for $\mu_n < U_n^0$,
which corresponds to the slab nuclei without the dripped neutrons.
For $\mu_n \geq U_n^0$,
\begin{eqnarray}
\tilde{n}_n^f &\equiv&
    \frac{2}{a N_k}\sum_{\alpha,k_z} \int \frac{dk_x dk_y}{(2\pi)^2}
    \theta(\mu_n-\epsilon_{\alpha,\bm{k}}^{(n)})
    \theta(\epsilon_{\alpha,\bm{k}}^{(n)}-U_n^0) \nonumber \\
    &=&
n_n - \frac{m_n}{\pi a N_k}\sum_{e_{\alpha,k_z}^{(n)}<U_n^0}
    \left(U_n^0-e_{\alpha,k_z}^{(n)} \right) .
\label{free_neutron_density_1}
\end{eqnarray}

The free neutrons of Eq.~(\ref{free_neutron_density_1}) includes those
with $\epsilon_{\alpha,\bm{k}}^{(n)}>U_n^0>e_{\alpha,k_z}^{(n)}$.
Since the Kohn-Sham Hamiltonian in the present case is separable in
each direction of $x$, $y$, and $z$,
the neutrons with $e_{\alpha,k_z}^{(n)}<U_n^0$
are practically bound in the $z$ direction.
In this sense, it may be reasonable to define
the ``free'' neutrons in the $z$ direction 
as those with $e_{\alpha,k_z}^{(n)} > U_n^0$.
\begin{eqnarray}
\bar{n}_n^f &\equiv&
    \frac{2}{a N_k}\sum_{\alpha,k_z} \int \frac{dk_x dk_y}{(2\pi)^2}
    \theta(\mu_n-\epsilon_{\alpha,\bm{k}}^{(n)})
    \theta(e_{\alpha,k_z}^{(n)}-U_n^0) \nonumber \\
    &=&
n_n - \frac{m_n}{\pi a N_k}\sum_{e_{\alpha,k_z}^{(n)}<U_n^0}
    \left(\mu_n-e_{\alpha,k_z}^{(n)} \right) .
\label{free_neutron_density_2}
\end{eqnarray}
Following these two definitions of ``free'' neutrons,
Eqs. (\ref{free_neutron_density_1}) and (\ref{free_neutron_density_2}),
we denote two kinds of effective mass as
\begin{equation}
\tilde{m}_z^* \equiv \frac{\tilde{n}_n^f}{\mathcal{K}^{zz}} 
	= m_n \frac{\tilde{n}_n^f}{n_z^c} ,
	\quad\quad
\bar{m}_z^* \equiv \frac{\bar{n}_n^f}{\mathcal{K}^{zz}} 
	= m_n \frac{\bar{n}_n^f}{n_z^c} ,
    \label{effective_mass_2}
\end{equation}
It is obvious that $n_n \geq \tilde{n}_n^f \geq \bar{n}_n^f$,
which leads to $m_z^* \geq \tilde{m}_z^* \geq \bar{m}_z^*$.

\section{Numerical results}
\label{sec:numerical}

The single-particle Hamiltonian in Eq. (\ref{sp_Hamiltonian_2})
has a property, $h_{k_z}^{(q)*}=h_{-k_z}^{(q)}$.
Therefore, the solutions of Eq. (\ref{Bloch_eq_2}) for negative $k_z$
can be constructed from those of positive $k_z$ as
$\phi_{-k_z}^{(q)}= \phi_{k_z}^{(q)*}$ with
$e_{\alpha,-k_z}^{(q)}=e_{\alpha,k_z}^{(q)}$.
The first Brillouin zone can be further reduced to $0\leq k_z \leq \pi/a$
for the present calculation.
When we solve Eq. (\ref{Bloch_eq_2}) for
discretized $k_z$ values of
\begin{equation}
    k_z=\frac{\pi}{a} \frac{l}{N_k'} ,
    \quad l=0,\cdots,N_k' ,
\end{equation}
the solutions for
$k_z=-\frac{\pi}{a} \frac{l}{N_k'}$ with $l=1,\cdots,N_k'-1$
are also obtained ($N_k'> 0$).
The number of $k_z$ points is $N_k=2N_k'$.
The calculation only with $k_z=0$ corresponds to $N_k=1$.

The $z$ coordinate in the unit cell ($0\leq z \leq a$)
is discretized in a mesh of $\Delta z=0.2$ fm,
and the nine-point finite-difference formula is used for
differentiation in Eq. (\ref{Bloch_eq_2}).
The number of iteration, shown in Sec.~\ref{sec:self-consistent_solutions},
necessary to reach the self-consistent solutions varies case by case;
a few tens to thousands of iteration.
Roughly speaking, more iterations are needed
for calculations of the slab phase at lower density.

\begin{figure}[tb]
\begin{centering}
\includegraphics[width=0.90\columnwidth]{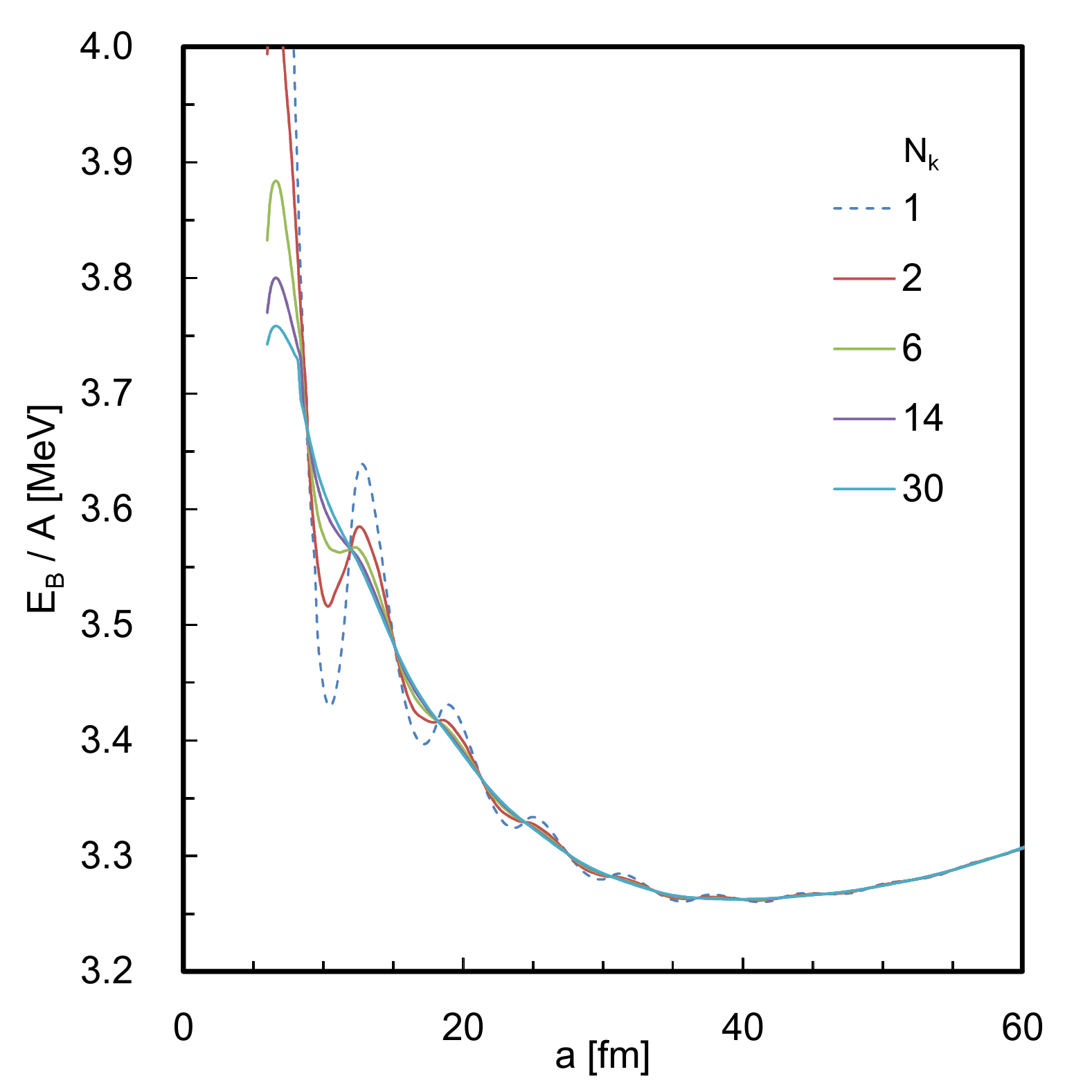}
\par\end{centering}
\caption{
(Color online)
Energy per nucleon $E_B/A$ as a function of lattice constant $a$,
for the case of proton fraction $Y_p=0.05$ and
baryon density $n_B=0.04$ fm$^{-1}$.
The blue solid line shows the one obtained with the band calculation
adopting $30$ points of the Bloch wave number,
while the blue dashed line indicates that of periodic boundary condition.
Those obtained with different $N_k$ are shown by solid lines with
different colors.
}
\label{fig:k-convergence}
\end{figure}

\subsection{Convergence with respect to number of $k$ points}
\label{sec:k-points}

First, we demonstrate that the periodic boundary condition,
which were often adopted in mean-field calculations for finite nuclei,
might lead to a wrong answer for the inner crust of neutron stars.
The simple periodic condition at the end of the unit cell,
$\varphi^{(q)}_{\alpha,\bm{k}}(x,y,z+a)=\varphi^{(q)}_{\alpha,\bm{k}}(x,y,z)$,
corresponds to the band calculation with $k_z=0$ only ($N_k=1$).
In Fig.~\ref{fig:k-convergence}, we show variation of energy
with respect to the lattice constant $a$
for the case of proton fraction $Y_p=0.05$ and
density $n_B=0.04$ fm$^{-1}$.
The result obtained with the periodic boundary condition shows
a strong oscillating pattern.
The optimum value of $a$ is given by
either $a=35.8$ fm or $41.4$ fm.

The calculation with periodic boundary condition may approximately provide
a proper answer for the outer crust.
However, for the inner crust,
because of the presence of the dripped neutrons at the boundary,
the energy shows a spurious oscillation as a function of $a$.
Since the calculation with $N_k$ $k_z$-points corresponds to
the one in a space of $0\leq z \leq N_k a$ with the periodic boundary condition,
this simple exercise indicates necessity of large-space calculation
for the inner crust.

This oscillation completely disappears for $a>10$ fm,
when we take into account
enough number of points for the Bloch wave number $k_z$.
In Fig.~\ref{fig:k-convergence}, we also show results with
different number of $N_k$.
Beyond $N_k=10$, we hardly see difference in the scale of 
Fig.~\ref{fig:k-convergence} for $a>10$ fm.
The converged result for the optimum value of the lattice constant
is $a=39.2$ fm.
In the present study, we adopt $N_k=30$ which is enough
to reach the convergence.
Although $N_k=30$ may not be sufficient
for a small lattice constant $a<10$ fm,
we have confirmed that the obtained optimum values of $a$ are
significantly larger than 10 fm.


\subsection{Thomas-Fermi calculation for slab phase}

In order to compare our result of the self-consistent band-theory calculation
with the one of the Thomas-Fermi (TF) theory,
we also perform the TF calculation.
Since the interaction energy of the BCPM functional is given
as a functional of $\rho_q(z)$,
we adopt exactly the same form in the TF calculation.
The kinetic density of Eq. (\ref{K_q}) is replaced by
\begin{equation}
\tau_q\left[\rho\right]=\frac{3}{5}\left(3\pi^2\right)^{2/3} \rho_q^{5/3}(z) 
    + \frac{\lambda_W}{\rho_q(z)} \left(\frac{d\rho_q}{dz}\right)^2 ,
    \label{TF_kinetic_density}
\end{equation}
where $\lambda_W$ is a parameter of
the Weizs{\"a}cker term 
given as $\lambda_W=1/36$ \cite{RS80}.

For practical solutions, we use the imaginary-time method
similar to Ref.~\cite{Lev84}.
Introducing the auxiliary functions $u_q(z)$ to represent the
density as $\rho_q(z)=u_q^2(z)$,
the time evolution is given by
\begin{equation}
\left.u_q(z)\right|_{t+\delta t}=u_q(z)-\delta t \left( h_q[\rho] -\mu_q \right) u_q(z) ,
\label{imaginary_time}
\end{equation}
where $u_q(z)$ and $\rho_q(z)$ in the right hand side are functions
at imaginary time $t$.
Here, the ``Hamiltonian'' $h_q[\rho]$ is given by
\begin{equation}
h_q\left[\rho\right] \equiv \frac{-1}{2m_q} 4\lambda_W \frac{d^2}{dz^2}
    + U_q(z)
    + \frac{1}{2m_q} \left\{ 3\pi^2 \rho_q(z) \right\}^{2/3} .
\label{h_TF}
\end{equation}
The ``chemical potentials'' $\mu_q$ are chosen to keep the average density
$n_q$ invariant for calculation with fixed $(n_B,Y_p)$.
\begin{equation}
    \mu_q = \frac{1}{a n_q}
    \int_0^a u_q(z) h_q[\rho] u_q(z) dz .
\end{equation}
Although $\mu_q$ conserves the average density in the first order in $\delta t$,
$u_q(z)$ are normalized every time to reproduce the given
average density $n_q$.
For calculation of beta equilibrium, $\mu_q$ are chosen to keep $n_B$ invariant,
which leads to
\begin{eqnarray}
\mu_q &=&  \frac{1}{n_B}
\Bigg[ \frac{1}{a}\sum_{q'=n,p}\int_0^a u_{q'}(z) h_{q'} u_{q'}(z) dz 
\nonumber \\
&&
+ n_{\bar q} \left\{ \left(m_{\bar q}-m_q\right)c^2
    + (-1)^{\delta_{qp}}\mu_e \right\}
\Bigg],
\end{eqnarray}
where $\bar q=n(p)$ for $q=p(n)$.

In practice, we adopt the imaginary-time step,
$\delta t=2^{-8}$ MeV$^{-1}$ for calculation with fixed $(n_B, Y_p)$,
and $\delta t=2^{-10}$ MeV$^{-1}$ for calculation of beta equilibrium.
The $z$ coordinate in the unit cell ($0\leq z \leq a$)
is discretized in a mesh of $\Delta z=0.2$ fm,
and the nine-point finite-difference formula is used for
differentiation in the imaginary-time evolution
of Eq.~(\ref{imaginary_time})
The convergence condition is given by
\begin{equation}
\frac{1}{n_B} \sum_q \int_0^a \left\{ (h_q - \mu_q)u_q(z) \right\}^2 dz
	< 10^{-12} \textrm{ MeV}^2.
\end{equation}
Since the TF theory directly treats the density instead of wave functions,
we may use the periodic boundary condition, $\rho_q(z)=\rho_q(z+a)$
which is equivalent to $u_q(z)=u_q(z+a)$ for
the real auxiliary function.

The last term in Eq. (\ref{TF_kinetic_density}) is called
the Weizs{\"a}cker correction term,
which takes into account the
inhomogeneity and has no contribution to the uniform matter.
The original value of $\lambda_W$, given by Weizs{\"a}cker, was
$1/4$ instead of $1/36$ \cite{Wei35}.
The factor $\lambda_W=1/36$ is consistent with the gradient expansion
and equivalently with the $\hbar$ expansion \cite{RS80}.
The importance of this Weizs{\"a}cker term will be demonstrated
in Sec.~\ref{sec:TF}.

\subsection{Comparison between band and TF calculations}
\label{sec:TF}

\begin{figure}[t]
\begin{center}
\includegraphics[width=0.90\columnwidth]{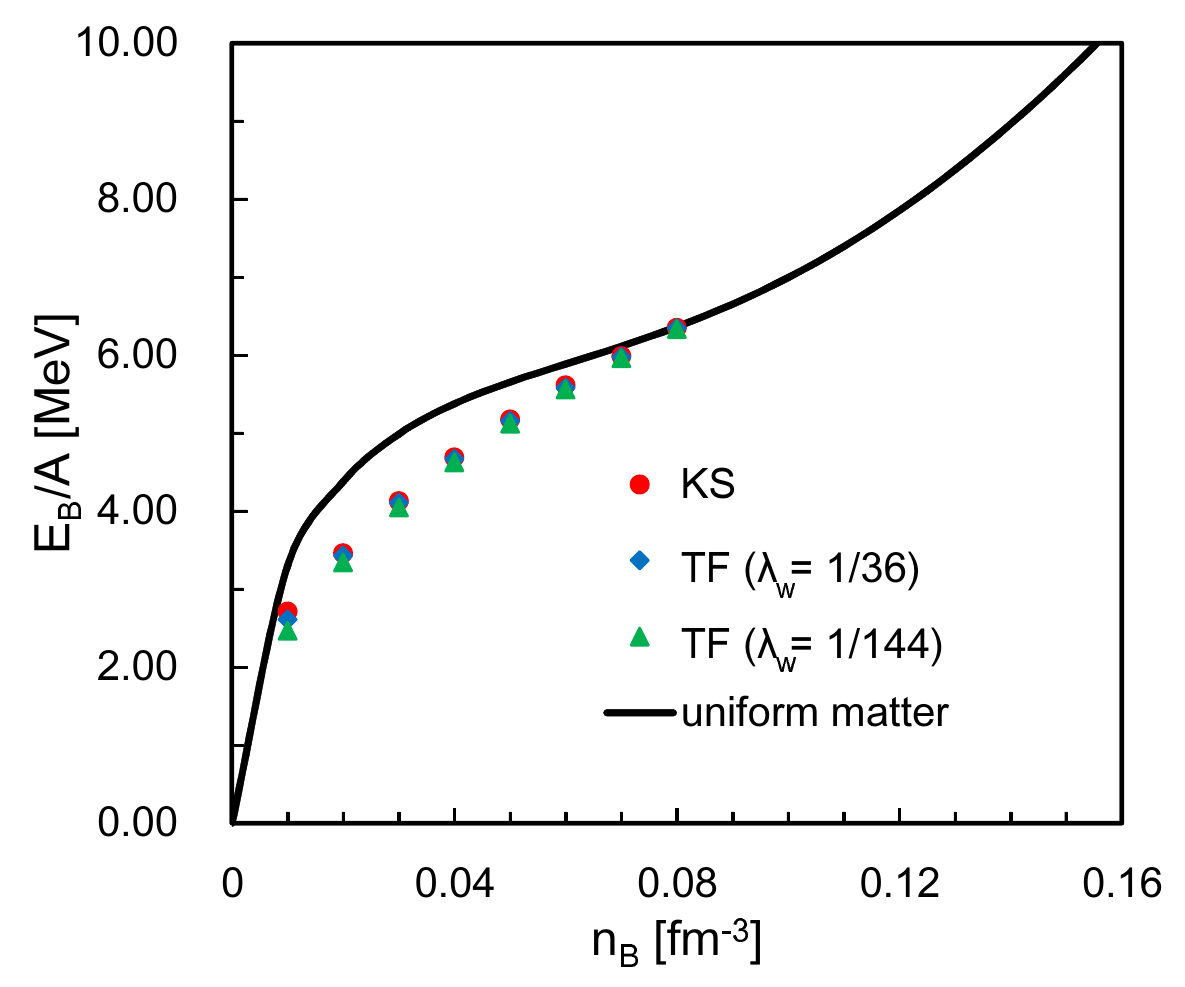}
\end{center}
\caption{
(Color online)
Calculated nuclear energy per nucleon $E_B/A$ at beta equilibrium
as a function of baryon density $n_B$.
The results of TF approximation are shown by
blue diamonds ($\lambda_W=1/36$)
and green triangles ($\lambda_W=1/144$),
while
those of the band calculation are by red circles.
The solid line represents $E_B/A$ for the uniform matter.
}
\label{fig:E_B/A_beta_equilibrium}
\end{figure}
\begin{figure}[h]
\begin{center}
\includegraphics[width=0.90\columnwidth]{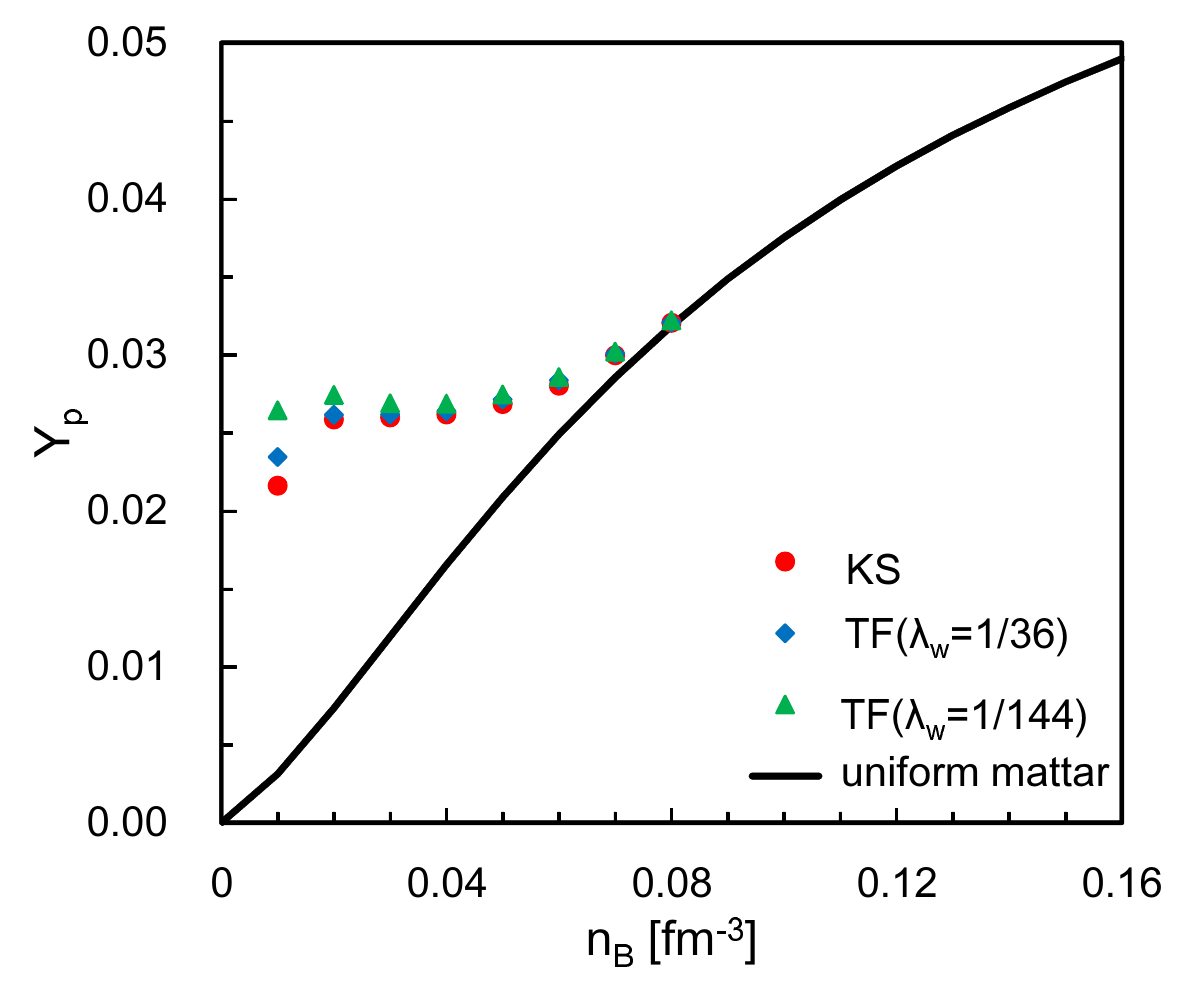}
\end{center}
\caption{
(Color online)
Calculated proton ratio $Y_p$ for slab phase as a function
of baryon density $n_B$.
See the caption of Fig.~\ref{fig:E_B/A_beta_equilibrium}.
}
\label{fig:Y_p_beta_equilibrium}
\end{figure}
\begin{figure}[h]
\begin{center}
\includegraphics[width=0.9\columnwidth]{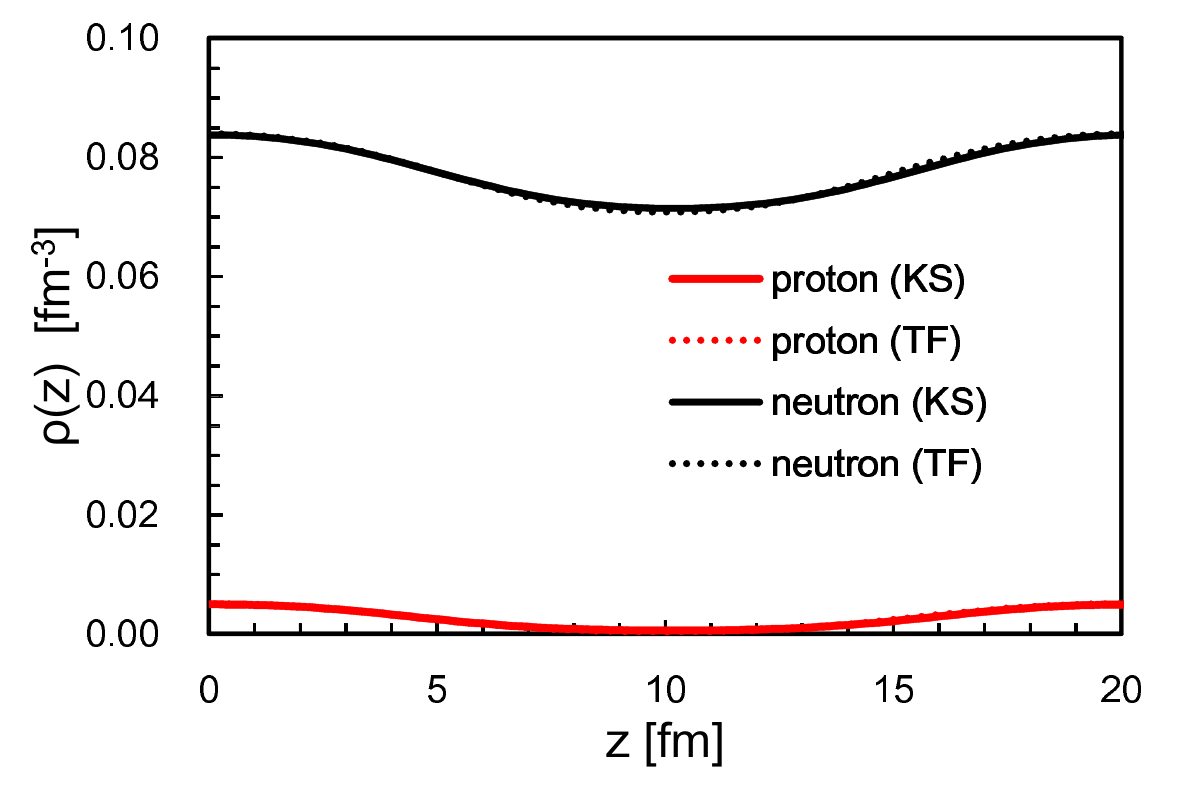}
\includegraphics[width=0.9\columnwidth]{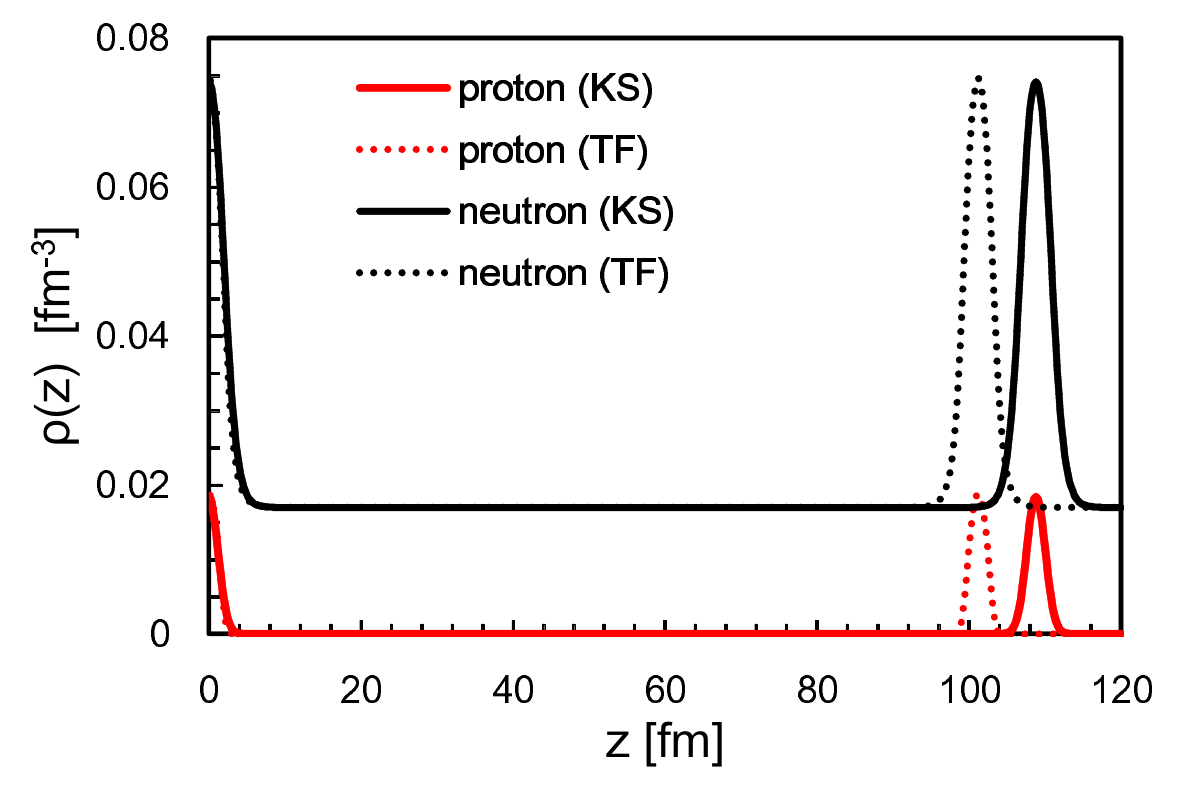}
\end{center}
\caption{
(Color online)
Calculated neutron and proton density distributions in the slab phase
as functions of $z$ at beta equilibrium.
The results of TF approximation ($\lambda_W=1/36$)
are shown by dashed lines.
The top panel shows that at $n_B=0.08$ fm$^{-3}$,
while the bottom one shows that at $n_B=0.02$ fm$^{-3}$.
The left end ($z=0$) corresponds to the center of a slab.
The obtained lattice constants are
$a=20.0$ fm for both the band and TF calculations
for the top panel, while they are
$a=108.8$ fm (band) and $101.4$ fm (TF)
for the bottom one.
}
\label{fig:density_profile_beta_equilibrium}
\end{figure}
\subsubsection{Beta equilibrium}

Let us first show comparison between the band calculation 
and the TF calculation at beta equilibrium.
We calculate the slab phase at baryon density every 0.01 fm$^{-3}$,
in a region of 
$0.01 \leq n_B \leq 0.08$.
At $n_B\geq (n_B)^c$ with the critical density
$(n_B)^c=0.083$ fm$^{-3}$, the slab nuclei are dissolved to produce
uniform nuclear matter in both calculations.
In Fig.~\ref{fig:E_B/A_beta_equilibrium}, 
we show the nuclear energy per nucleon $E_B/A$ which
monotonically increases as
the baryon density $n_B$ increases;
$E_B/A = + 2.7 $ MeV at $n_B=0.01$ fm$^{-3}$ and
$E_B/A = + 6.3$ MeV at $n_B=0.08$ fm$^{-3}$.
The energy gain from the uniform phase amounts to
$E_B/A\approx 1$ MeV at maximum.
The TF calculation well reproduces these values,
especially near the critical density $n_B\sim (n_B)^c$.

The calculated proton ratio $Y_p$ is shown in
Fig.~\ref{fig:Y_p_beta_equilibrium}.
$Y_p$ at beta equilibrium in the slab phase is
around $Y_p\approx 0.02-0.03$.
Because of this small value of $Y_p$, the neutrons are always dripped
in the slab phase at beta equilibrium.
The band calculation suggests monotonic increase of $Y_p$
as a function of density,
and it reaches $Y_p=0.032$ at $n_B=0.08$ fm$^{-3}$.
In the low-density side,
it shows an approximate constant value of
$Y_p\approx 0.026$ in a region of $0.02\leq n_B \leq 0.04$ fm$^{-3}$,
and a sudden drop from $Y_p\approx 0.026$ at $n_B=0.02$ to 
$Y_p\approx 0.022$ at $n_B=0.01$ fm$^{-3}$.
The TF calculation gives slightly larger $Y_p$ values,
which is more prominent at lower density.

Figure~\ref{fig:density_profile_beta_equilibrium} shows
the one-dimensional density profiles.
Qualitative features of the density profiles are well reproduced 
in the TF calculation,
though we find some quantitative difference, especially at low density.
The TF calculation always predicts the lattice constant $a$ smaller than
the band calculation.
This is probably due to the fact that the TF approximation
underestimates the surface diffuseness.
The TF calculation \cite{Oya93} predicted that the slab phase appears
near the bottom of the
inner crust of neutron star, at density around $n_B=0.07-0.09$ fm$^{-3}$.
In such density region, since the difference is not so large,
the TF description of the slab phase is reasonably good.
In contrast, at lower density, the difference becomes larger.
For instance, at $n_B=0.01$ fm$^{-3}$, the lattice constant of $a=199.6$ fm
is predicted by the band calculation,
while $a=176.2$ fm by the TF calculation.

In the uniform limit, the TF calculation exactly reproduces the result
of the band calculation.
Thus, it is natural to observe that the TF calculation well agrees with
the band calculation at the uniform limit,
$n_B \rightarrow 0.09$ fm$^{-3}$.
The quantum effect missing in the TF calculation is more important
at low baryon density $n_B$.

In order to see the effect of the Weizs{\"a}cker correction term
in the TF calculation,
we perform the same calculation with the prefactor
reduced by $1/4$, $\lambda_W=1/144$.
The energy $E_B/A$ is not sensitive to this change, however,
the density profiles, the proton ratio $Y_p$, and the optimal slab
interval $a$ are affected by the Weizs{\"a}cker term.
The calculated proton ratio $Y_p$ are shown by triangles
in Fig.~\ref{fig:Y_p_beta_equilibrium}.
The deviation is larger at lower density.

Another feature we find in the band calculation is that
not only neutrons but also protons are dripped from slab nuclei
at $n_B\geq 0.08$ fm$^{-3}$.
Near the transition to the uniform matter, the nuclear potential
becomes almost flat.
Thus, although protons are deeply bound with the chemical
potential $\mu_p \approx -70$ MeV,
$\mu_p$ can be larger than the maximum potential value for protons, $U_p^0$.
This proton drip phenomenon does not take place in the TF
calculation.

\subsubsection{Fixed proton ratio}

\begin{figure}[t]
\begin{centering}
\includegraphics[width=0.88\columnwidth]{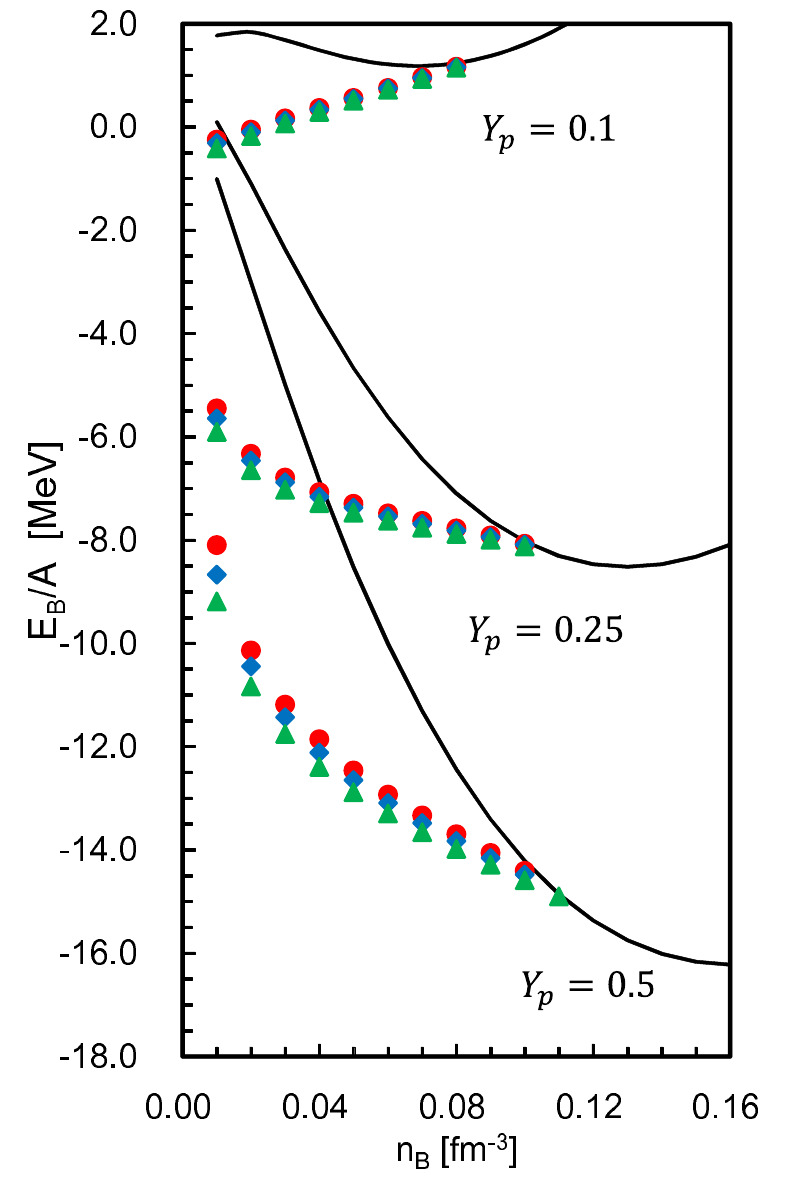}
\par\end{centering}
\caption{
(Color online)
Calculated nuclear energy per nucleon $E_B/A$ as functions of
baryon density $n_B$, for the proton ratio $Y_p=0.5$,
$Y_p=0.25$, and $Y_p=0.1$.
The results of the TF calculation are shown by blue diamonds
($\lambda_W=1/36$) and by green triangles ($\lambda_W=1/144$), while
those of the band calculation are by red circles.
The solid lines show those of the uniform matter.
}
\label{fig:E/A_fixed_Y_p}
\vspace{5pt}
\begin{centering}
\includegraphics[width=0.8\columnwidth]{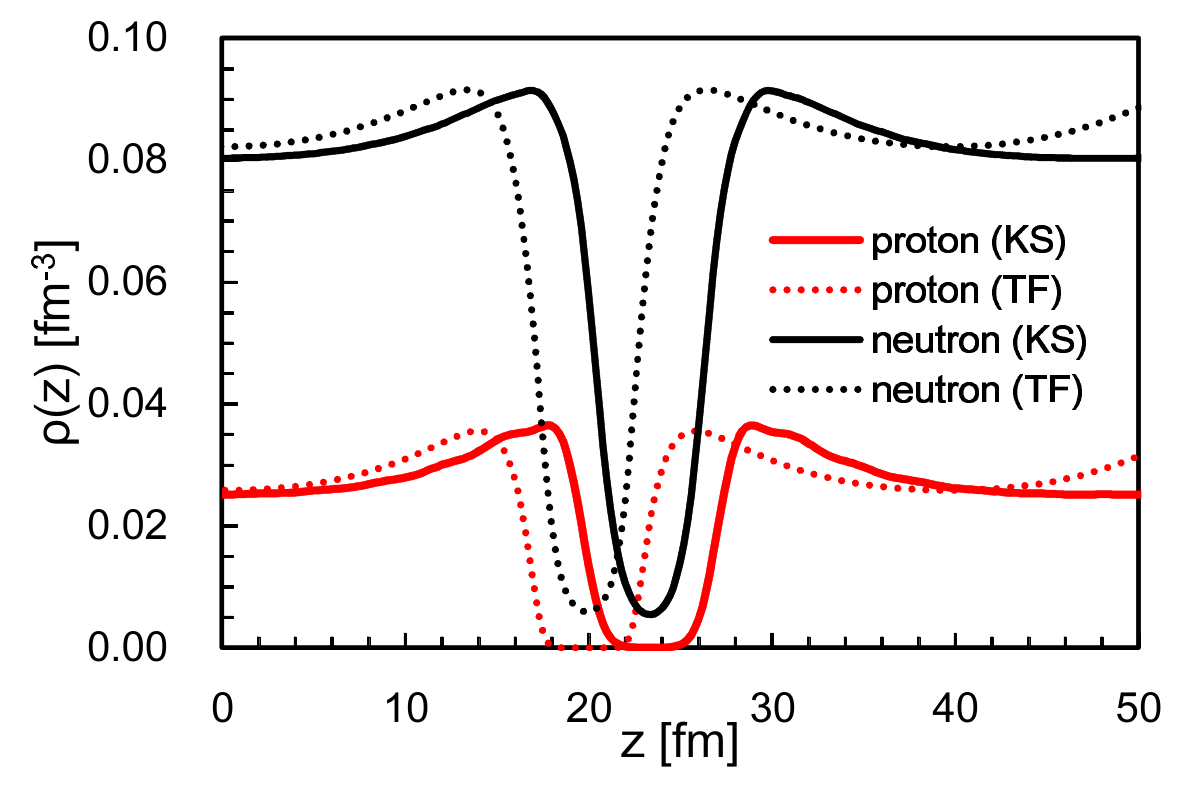}
\par\end{centering}
\caption{
(Color online)
The same as Fig.~\ref{fig:density_profile_beta_equilibrium},
but at $n_B=0.1$ fm$^{-3}$ and $Y_p=0.25$.
}
\label{fig:density_profile_fixed_Y_p}
\end{figure}

Away from the beta equilibrium, 
the discrepancy between the band and TF calculations
becomes more prominent,
especially when the lattice constant $a$ is large.
We show, in Fig.~\ref{fig:E/A_fixed_Y_p},
$E_B/A$ as functions of density $n_B$
for different values of proton ratio $Y_p$.
At low density ($n_B=0.01$ fm$^{-3}$),
the TF calculation underestimates $E_B/A$ by
$\Delta(E_B/A)=0.58$ MeV
for the symmetric slab case with $Y_p=0.5$.
Adopting the reduced value of $\lambda_W=1/144$, 
the difference in $E_B/A$ becomes even larger,
which amounts to 1.09 MeV.
This discrepancy becomes smaller at higher density, however,
the structure of the slab nuclei can be very different.
As an example,
we show the nucleons' density profiles
near the transition point from slab to uniform phase,
in Fig.~\ref{fig:density_profile_fixed_Y_p}.
The structure at $n_B=0.1$ fm$^{-3}$ with $Y_p=0.25$
is like ``anti-slab'' phase where gaps periodically appear in
the uniform matter.
Although the shapes of density distribution are somewhat similar,
the lattice constant of the TF calculation is very different from
that of the band calculation.
In Fig.~\ref{fig:density_profile_fixed_Y_p},
the optimal lattice constant is $a=49.4$ fm
for the band calculation,
while $a=39.6$ fm for the TF calculation.
A depression of density at the center of each slab,
which appears for relatively high $Y_p$ values,
is due to the Coulomb interaction among protons.

Increasing the baryon density, the slab phase changes into
the uniform phase at a critical value of density.
The band calculation indicates that
this critical density $(n_B)^c$ is located at
between 0.1 and 0.11 fm$^{-3}$ for $Y_p=0.5$,
and
between 0.08 and 0.09 fm$^{-3}$ for $Y_p=0.1$.
The quantum fluctuation tends to make the density distribution flatter,
which leads to lower values of $(n_B)^c$ than those of the TF approximation.

\begin{figure}[t]
\begin{centering}
\includegraphics[width=0.80\columnwidth]{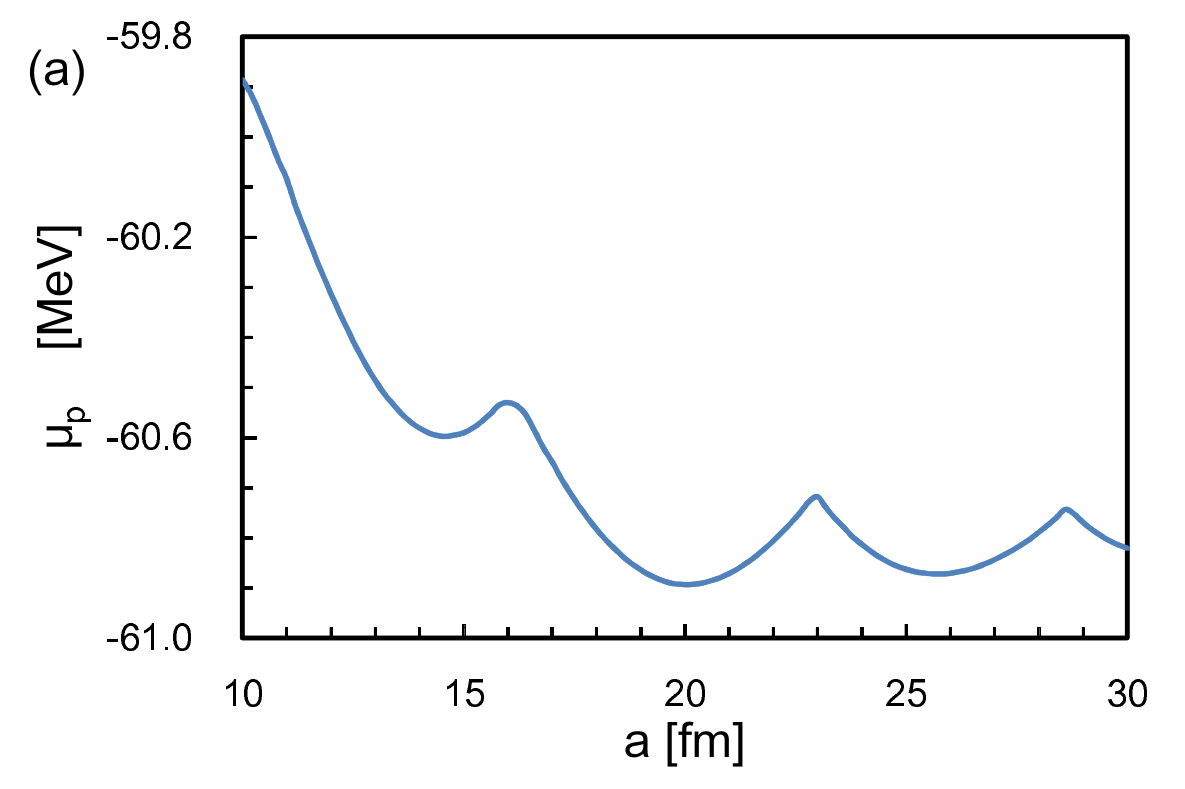}
\includegraphics[width=0.80\columnwidth]{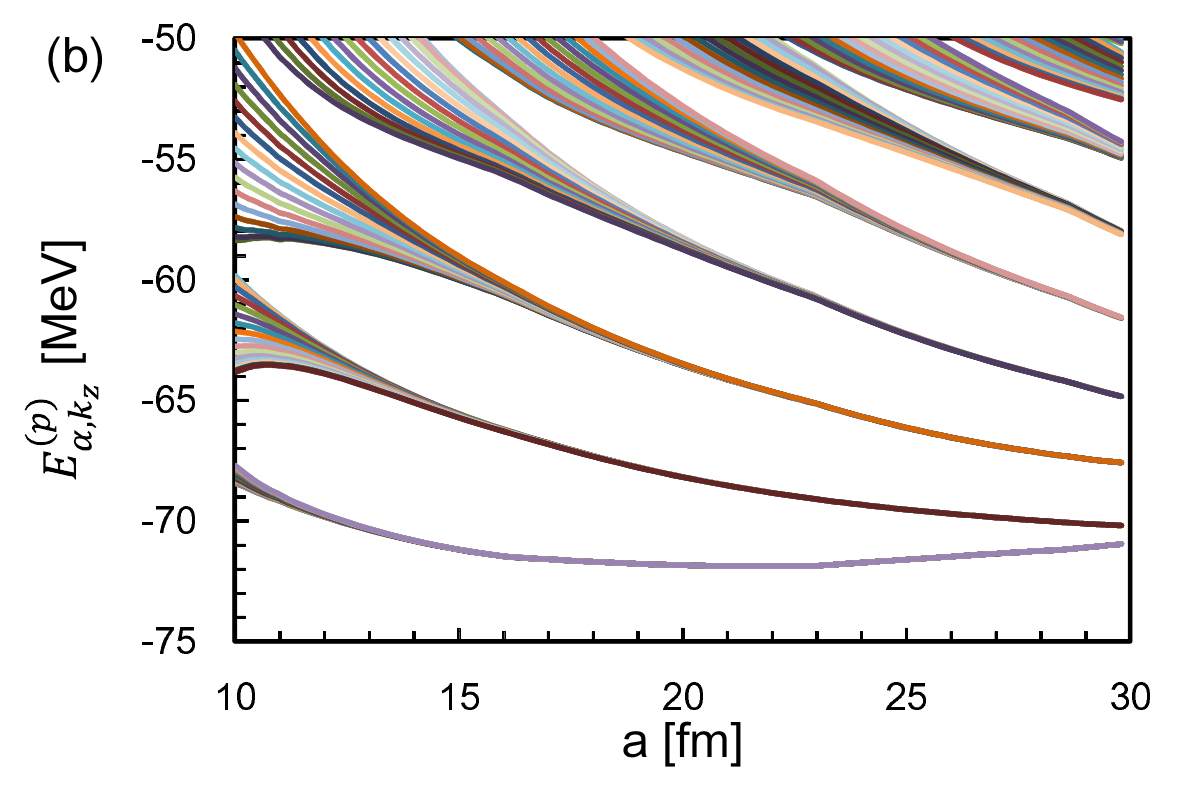}
\includegraphics[width=0.80\columnwidth]{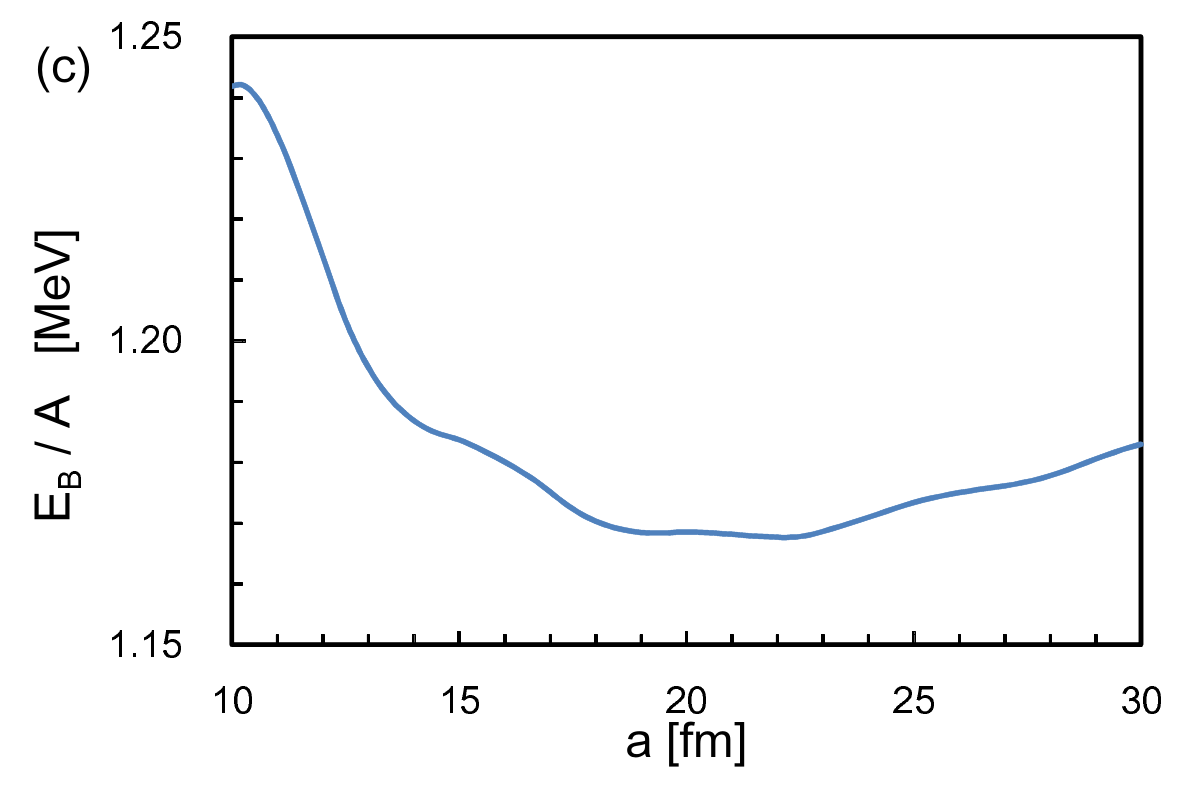}
\par\end{centering}
\caption{
(Color online)
(a) Calculated chemical potential for protons,
(b) single-particle energies for protons,
and (c) nuclear energy per baryon,
as functions of lattice constant $a$,
for the case of $Y_p=0.1$ and $n_B=0.08$ fm$^{-3}$.
In the panel (b), each bundle of lines, which correspond to a single band
index $\alpha$, contains 16 lines with different Bloch wave numbers $k_z$.
}
\label{fig:proton_shell_effect}
\end{figure}

\subsection{Band structure for slab phase}

\subsubsection{Proton shell effect}

The shell effect is an obvious missing piece in the TF approximation.
In the one-dimensional slab phase, 
Since all the nucleons have two-dimensional free motion in the
$x-y$ directions,
the shell effect is not so strong.
Nevertheless, in most cases, the protons are bound in the $z$ direction
and we expect some shell effect due to change of protons' orbital occupancy.

In Fig.~\ref{fig:proton_shell_effect} (a), we show an example
of the proton chemical potential $\mu_p$
as a function of the lattice constant $a$.
It clearly shows ridges and multiple minima which are associated with the
proton shells.
The corresponding single-particle (band) energies $e^{(p)}_{\alpha,k_z}$
are shown in Fig.~\ref{fig:proton_shell_effect} (b).
At the crossing points between the single-particle energy and
the chemical potential, the chemical potential shows cusps.
Although the proton shell effect is clearly visible in the chemical
potential, the energy per baryon, $E_B/A$, still represents
a smooth curve as in Fig.~\ref{fig:proton_shell_effect} (c).
At the crossing points, we observe a kind of bending of the curve,
with different slopes before and after the crossing.
It should be noted that the proton single-particle energies
$\epsilon_{\alpha,\bf{k}}$ in Eq.~(\ref{sp_energy})
are not discrete, because $k_\rho^2/2m_p$ is continuous.

The energies $e^{(p)}_{\alpha,k_z}$ should not depend on $k_z$
for bound protons.
Therefore, the unraveled bundle of lines corresponds to dripped orbitals.
Figure~\ref{fig:proton_shell_effect} (b) indicates
that the protons start to drip from the slab nuclei at small $a$.

\subsubsection{Band structure of neutrons}

\begin{figure*}[tb]
\begin{centering}
\includegraphics[width=0.85\columnwidth]{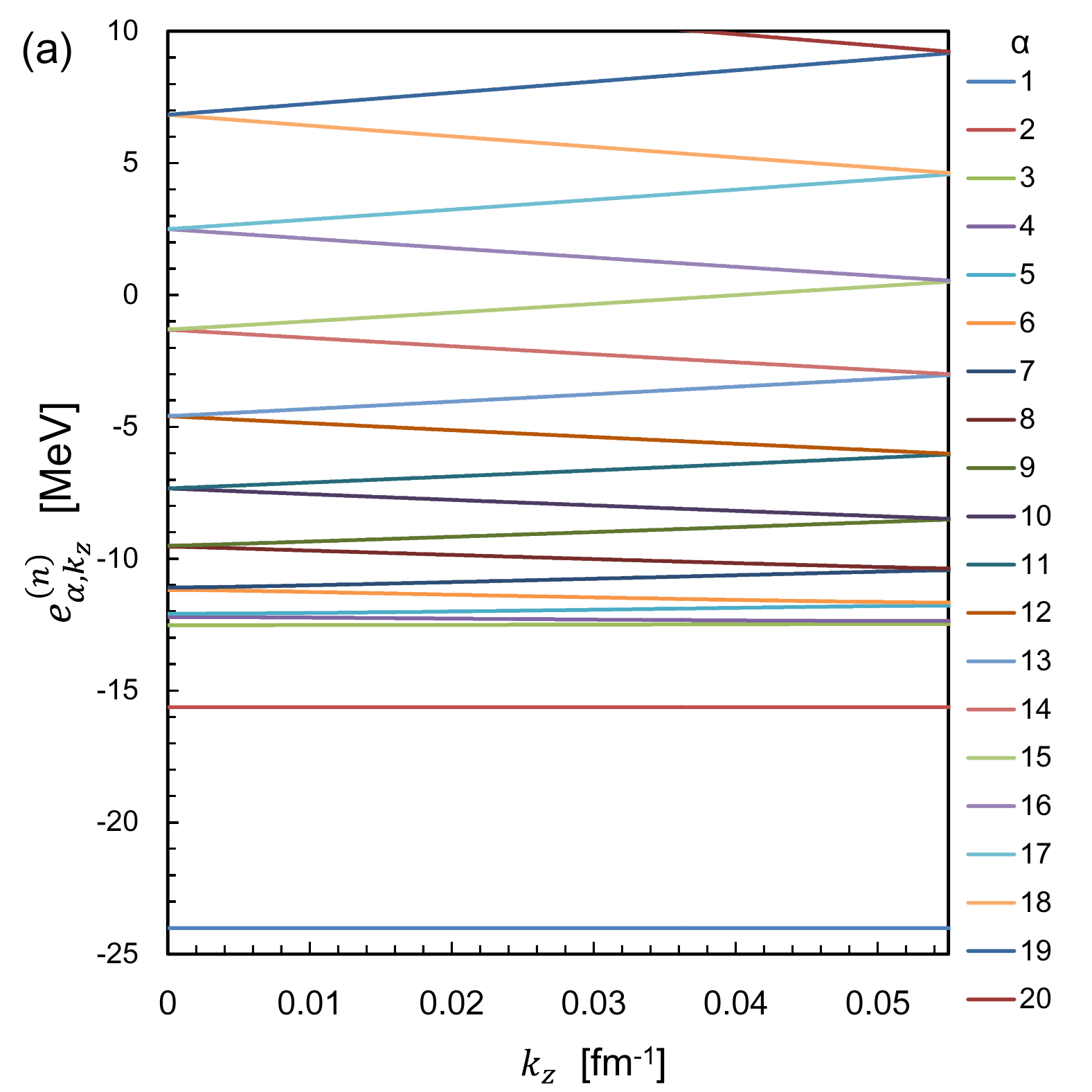}
\includegraphics[width=0.85\columnwidth]{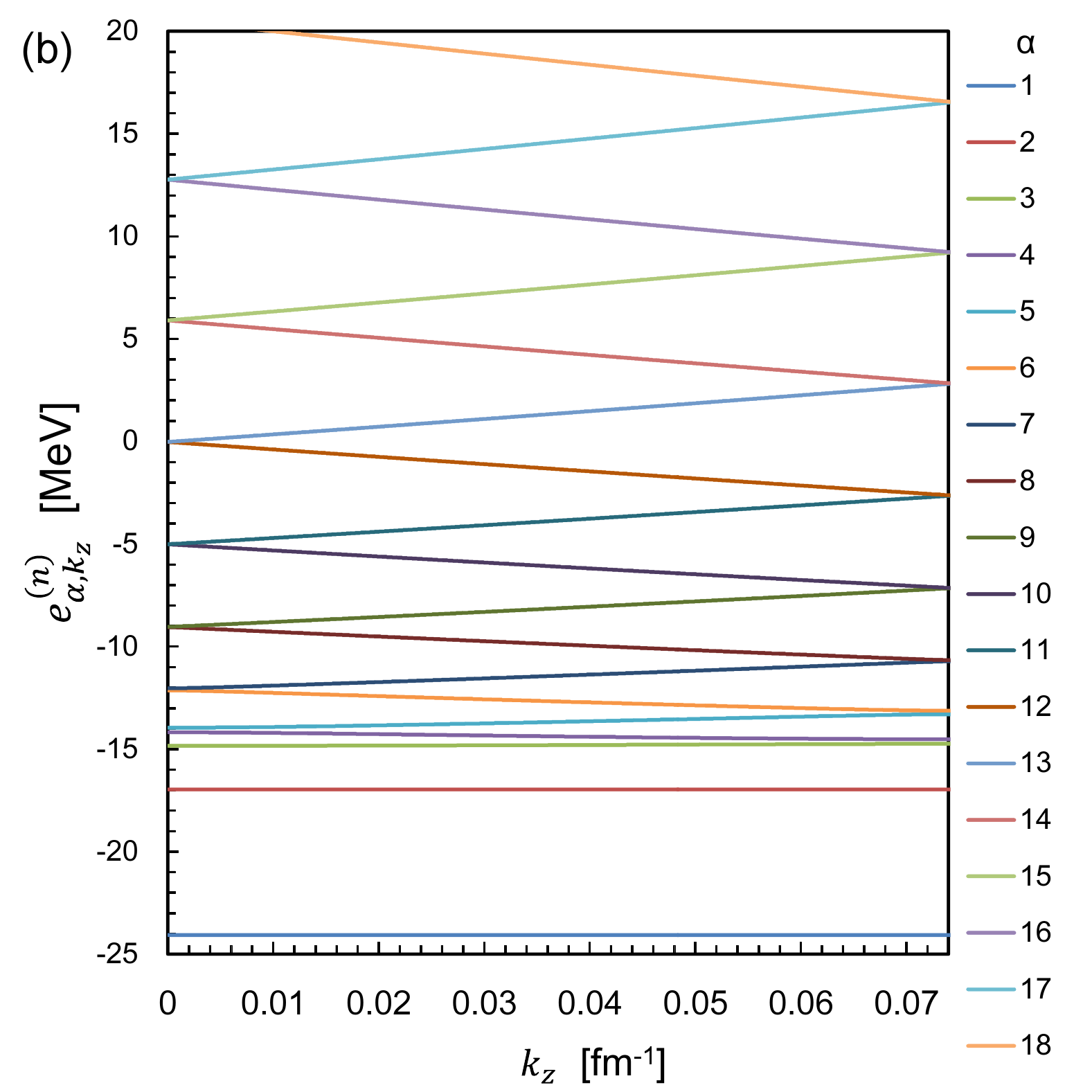}
\includegraphics[width=0.85\columnwidth]{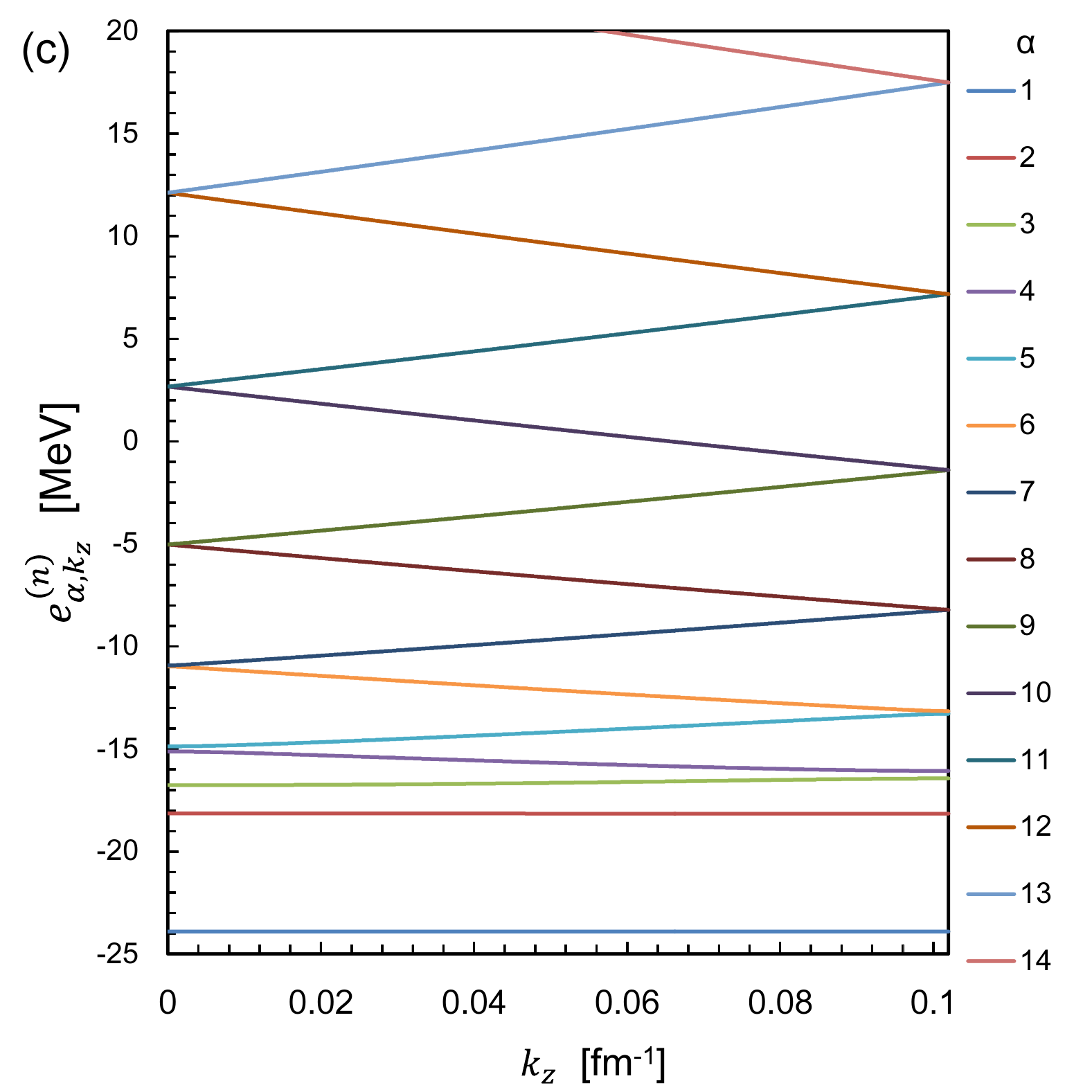}
\includegraphics[width=0.85\columnwidth]{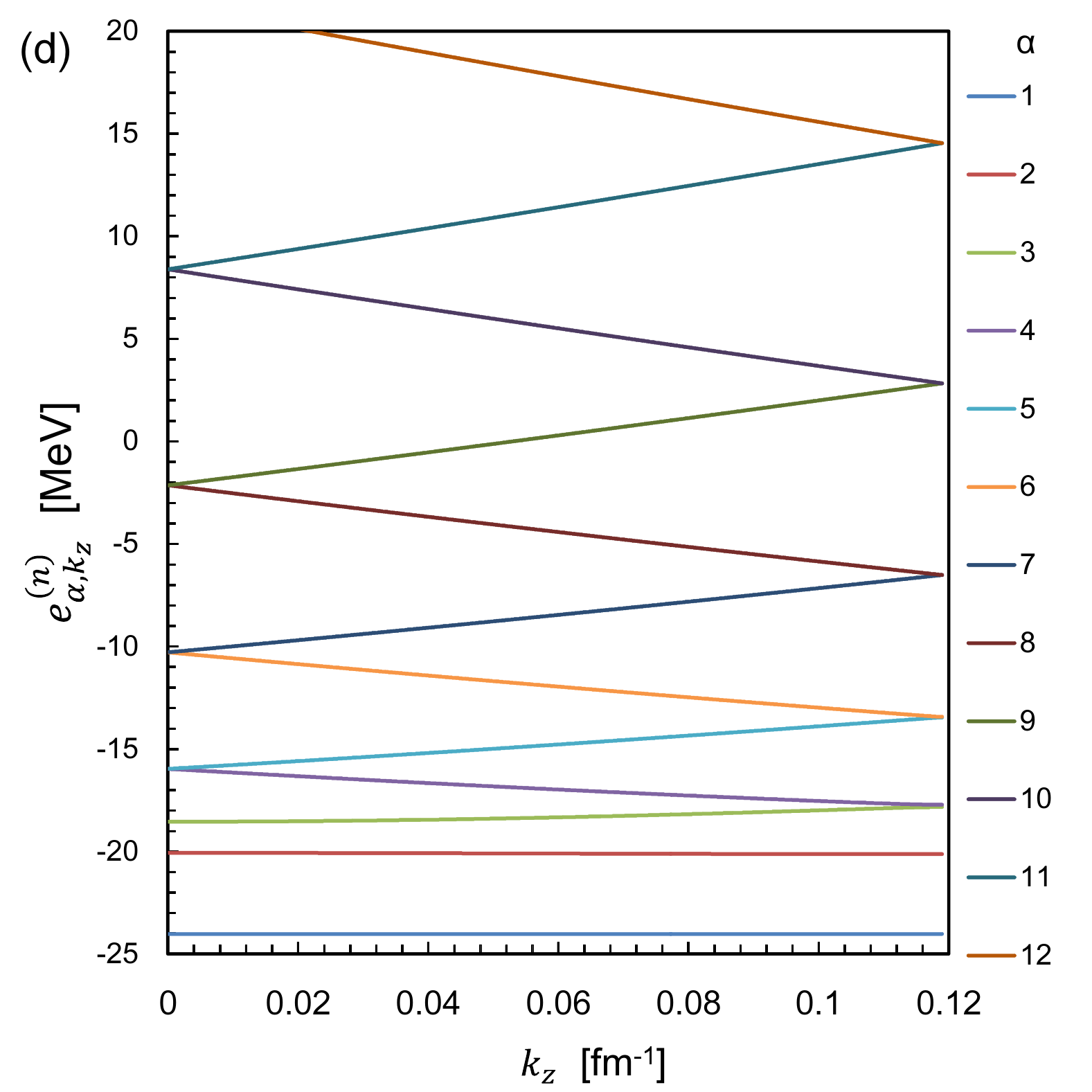}
\par\end{centering}
\caption{
(Color online)
Neutron band structure at beta equilibrium
as a function of Bloch wave number $k_z>0$;
(a) $n_B=0.04$ fm$^{-3}$,
(b) $n_B=0.05$ fm$^{-3}$,
(c) $n_B=0.06$ fm$^{-3}$, and
(d) $n_B=0.07$ fm$^{-3}$.
The lattice constants and neutron chemical potentials are given as
(a) $a=56.4$ fm, $\mu_n=8.45$ MeV, (b) 42.2 fm, 9.55 MeV, (c) 30.8 fm, 10.71 MeV,
(d) 26.4 fm, and 11.93 MeV,
respectively.
In each panel, the right end corresponds to the end point of the
Brillouin zone $k_z=\pi/a$.
}
\label{fig:neutron_band}
\end{figure*}

In the inner crust of neutron stars, the neutrons are partially dripped
from the slabs.
In Fig.~\ref{fig:neutron_band}, we show the band structure
of neutrons at beta equilibrium.
Since the band structure is symmetric with respect
to the transformation of $k_z\rightarrow -k_z$,
we show here only a half of the first Brillouin zone with
positive $k_z$.
At the center ($k_z=0$) and the end ($k_z=\pm \pi/a$) of first Brillouin zone,
there are band gaps.
The calculated band gaps are small in the slab phase,
which are less than a few hundreds of keV.
The magnitude of the band gap varies from band to band,
however, typically it becomes smaller for bands with larger energies
$e^{(n)}_{\alpha,k_z}$.

\begin{figure}[htb]
\begin{centering}
\includegraphics[width=0.90\columnwidth]{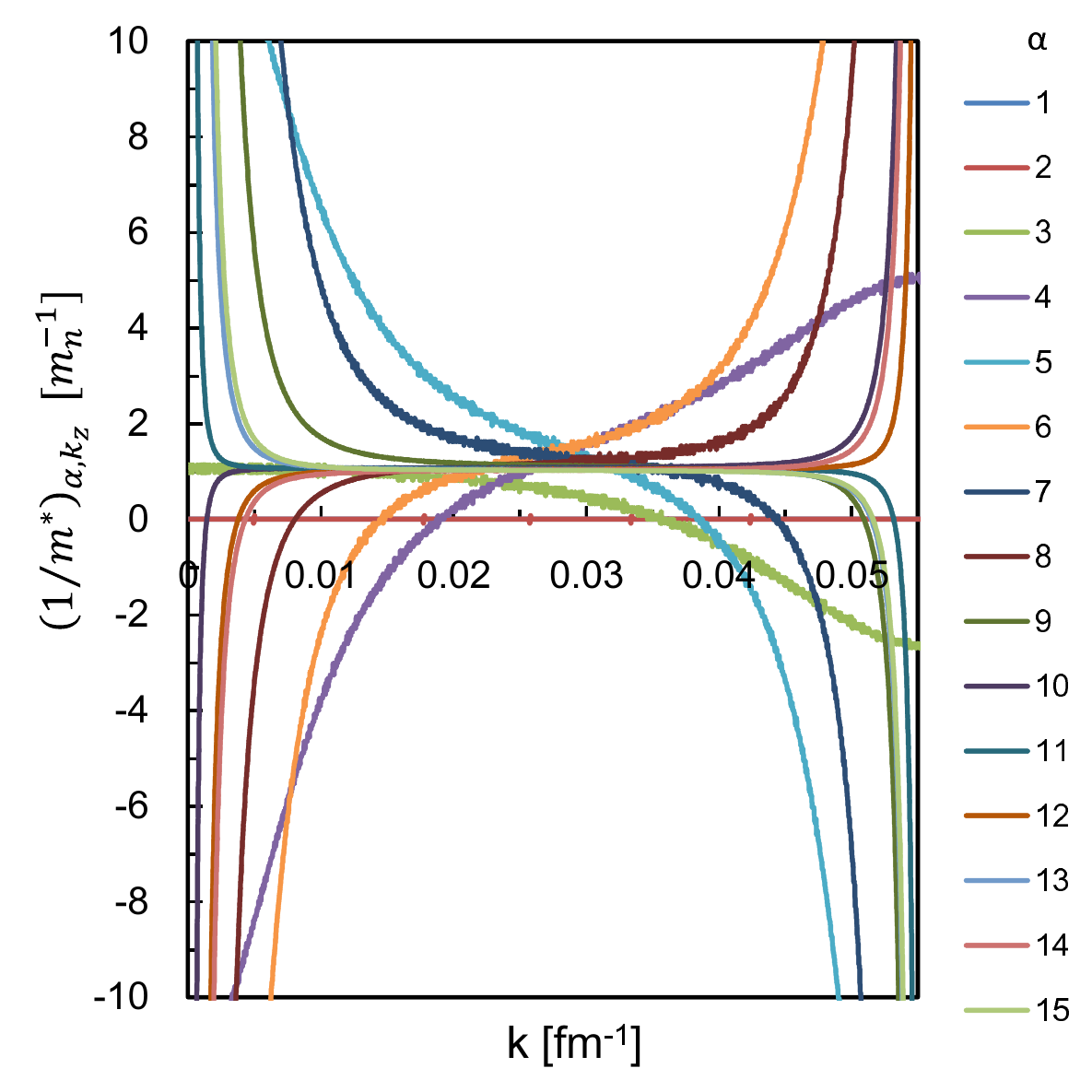}
\par\end{centering}
\caption{
(Color online)
Neutron effective mass $(1/m^*)_{\alpha,k_z}^{zz}$ as a function of $k_z$,
calculated at beta equilibrium with $n_B=0.04$ fm$^{-3}$.
The colors correspond to those in Fig.~\ref{fig:neutron_band}(a).
}
\label{fig:effective_mass_t}
\end{figure}

The effective mass (\ref{effective_mass_t}) along the $z$ direction
is shown in Fig.~\ref{fig:effective_mass_t} for $n_B=0.04$ fm$^{-3}$.
For bound neutrons with $e^{(n)}_{\alpha,k_z} < U_n^0$,
we have $(1/m_z^*)_{\alpha,k_z}^{zz}\approx 0$ which means that
these neutrons have infinite mass, because they cannot ``move''
toward $z$ direction.
Here, since we have $U_n^0=-12.57$ MeV,
the lowest two bands correspond to these ``core'' neutrons
with the infinite effective mass.
This is also consistent with vanishing group velocity 
according to Eq. (\ref{group_velocity}).

The dripped neutrons have finite values of effective mass.
The lowest ``valence'' band ($e^{(n)}_{\alpha,k_z} > U_n^0$)
is the third lowest one in Fig.~\ref{fig:neutron_band}(a).
In Fig.~\ref{fig:effective_mass_t}),
only this band shows the convergence of the effective mass
to the bare mass
($1/m_n=1.06 \times 10^{-3}$ MeV$^{-1}$)
at $k_z\rightarrow 0$,
while it becomes negative at $|k_z|> 0.036$ fm$^{-1}$.
If we apply a force on neutrons toward positive $z$ direction,
those with negative effective mass would be accelerated
toward negative $z$ direction.
This is due to the Bragg scattering from the periodic potential.
For high-lying bands, only at the vicinity of $k_z=0$ and $\pm \pi/a$,
the effective mass shows deviation from the bare mass.

In solids with three-dimensional crystalline structure,
the electrons in fully occupied bands
do not contribute to the conduction current
\cite{AM76}.
This is because all the states in the band $\alpha$ are occupied,
thus, the intraband excitations,
$(\alpha,\mathbf{k})\rightarrow(\alpha,\mathbf{k}')$,
are not allowed.
However, the situation is different in the 
slab phase embedded in the three-dimensional matter.
Even for $e^{(n)}_{\alpha,k_z}<\mu_n$,
there are still unoccupied states with the same $(\alpha,k_z)$,
because it is always possible to find the states with
$\epsilon^{(n)}_{\alpha,\mathbf{k}}>\mu_n$ by increasing ${k}_\rho$,
according to Eq. (\ref{sp_energy}).
Therefore, all the ``valence'' bands with finite velocity
of the $z$ direction,
$de^{(n)}_{\alpha,k_z}/dk_z\neq 0$,
may contribute to the neutron current.

\begin{figure}[tb]
\begin{centering}
\includegraphics[width=0.90\columnwidth]{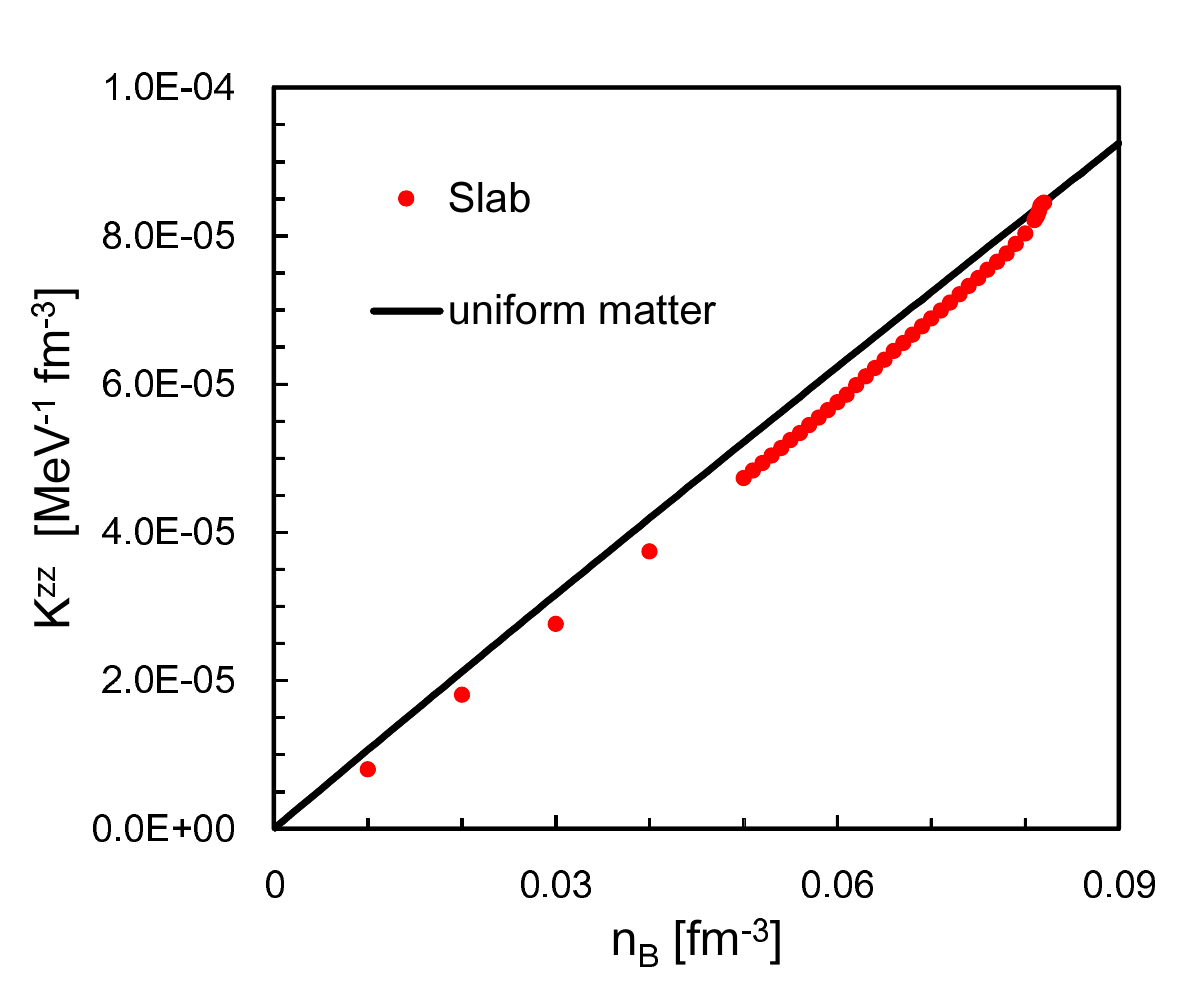}
\par\end{centering}
\caption{
(Color online)
Neutron mobility coefficient $\mathcal{K}^{zz}$
calculated at beta equilibrium as a function of density.
Those of the uniform matter are shown by a solid line.
}
\label{fig:mobility}
\end{figure}
\begin{figure}[htb]
\begin{centering}
\includegraphics[width=0.90\columnwidth]{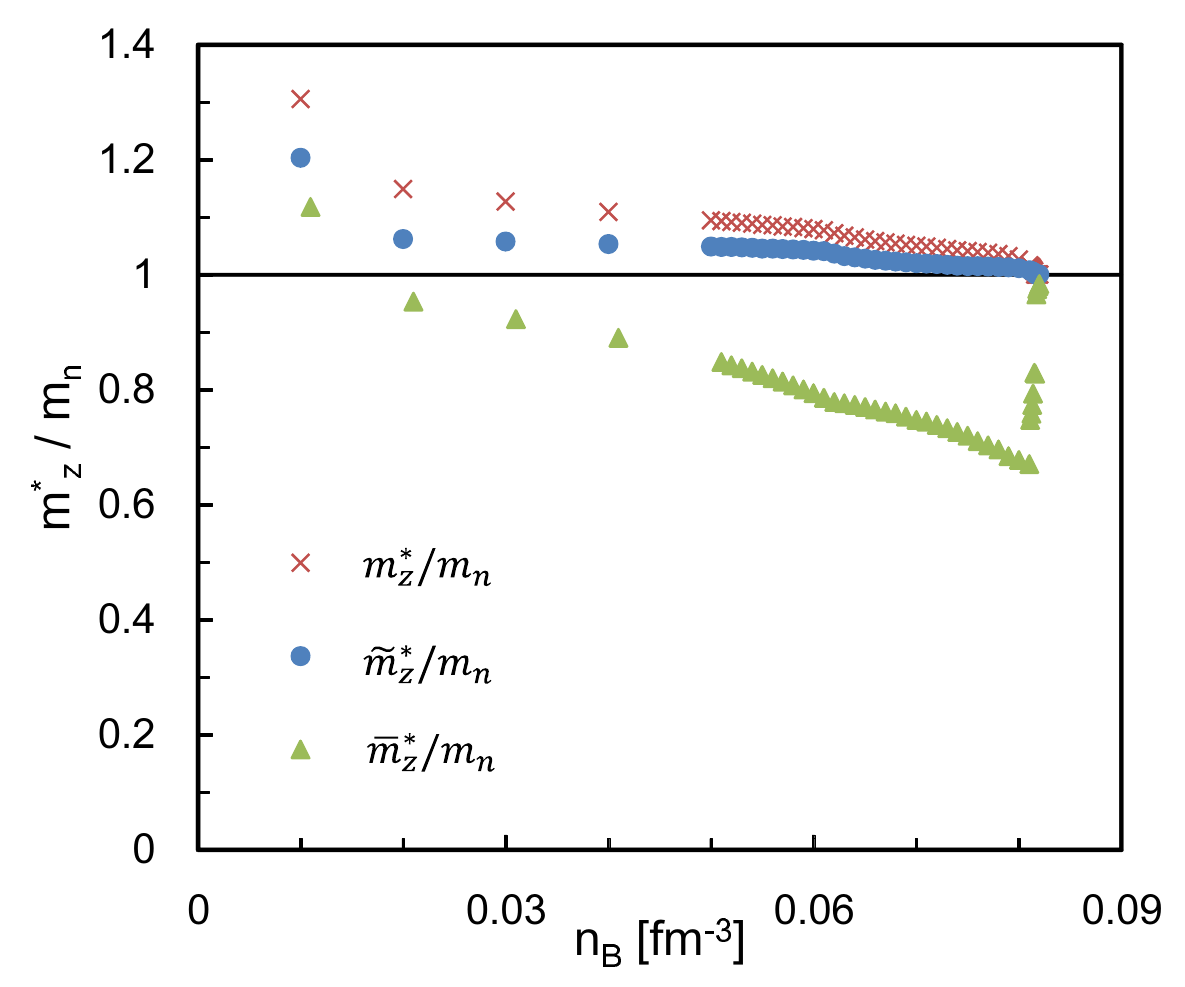}
\par\end{centering}
\caption{
(Color online)
Neutron effective masses $m_z^*$, $\bar{m}_z^*$, and $\tilde{m}_z^*$
calculated at beta equilibrium
as functions of density.
}
\label{fig:effective_mass}
\end{figure}

\subsubsection{Mobility and average effective mass of neutrons}

In the uniform matter, the mobility coefficient (\ref{mobility_coefficient_z})
for neutrons is simply given as
$\mathcal{K}^{zz}=n_z^c/m_n=n_n/m_n$,
which is proportional to the total neutron density $n_n$.
Since the proton ratio $Y_p$ is approximately constant at beta equilibrium,
this simple relation leads to
the dashed straight line in Fig.~\ref{fig:mobility}.
In the slab phase, the mobility toward the $z$ direction
is certainly reduced from that in the uniform matter.
This directly affects the average effective mass $m_z^*$ of
Eq. (\ref{effective_mass_1}),
which leads to $m_z^*/m_n\approx 1.0-1.3$,
as a monotonically decreasing function of density $n_B$
(crosses in Fig.~\ref{fig:effective_mass}).
A major origin of this effect comes from the existence of bound neutrons
which are practically prevented from moving in the $z$ direction
by the potential barrier, $e^{(n)}_{\alpha,k_z}<U_n^0$.
These ``core'' neutrons exist inside the slab nuclei but
are not present in the dripped low-density neutrons between the slabs.

In order to measure the mobility of these low-density neutrons
dripped from the slab nuclei,
we should calculate the density of ``free'' neutrons $n_n^f$ instead of
the total neutron density $n_n$.
According to the definition, Eqs. (\ref{free_neutron_density_1}),
and (\ref{free_neutron_density_2}),
$n_n^f$ are calculated.
At beta equilibrium with $n_B\geq 0.83$ fm$^{-3}$, 
the calculation predicts the uniform phase
in which we have $n_n^f=n_n$ because $U_n^0 < e^{(n)}_{\alpha,k_z}$
for all $(\alpha,k_z)$.
$n_n^f$ monotonically decreases as decreasing $n_B$,
then, reaches $\tilde{n}_n^f/n_n= 0.922$
and $\bar{n}_n^f/n_n= 0.856$ at $n_B=0.01$ fm$^{-3}$.

We show in Fig.~\ref{fig:effective_mass}
the average effective mass, $m_z^*$, $\tilde{m}_z^*$, and $\bar{m}_z^*$.
For their definitions, see Eqs. (\ref{effective_mass_1}) and
(\ref{effective_mass_2}).
It turns out that the average effective mass depends on the definition
of the free neutrons.
All the three effective masses are larger than the bare neutron mass
$m_n$ at very low density ($n_B\lesssim 0.01$ fm$^{^3}$).
However, we have $\bar{m}_z^*/m_n<1$ at $n_B\gtrsim 0.02$ fm$^{-3}$,
thus, $m_z^* \geq \tilde{m}_z^* \geq m_n \geq \bar{m}_z^*$
in the region of $0.02<n_B<0.083$ fm$^{-3}$.
Beyond the critical density $n_B>0.083$ fm$^{-3}$,
all the effective masses are identical to the bare mass.
Since we expect that the dripped neutrons between the slab nuclei
consist of neutrons occupying states with $e^{(n)}_{\alpha,k_z}>U_n^0$,
we suppose that $\bar{m}_z^*$ represents the
mobility of the dripped low-density neutrons.
They are calculated to be smaller than the bare mass.
This means that the conduction neutron density is larger than that of
free neutrons, $n_z^c > \bar{n}_n^f$.
Thus, the entrainment effect for the dripped neutrons in the slab phase
enhances the mobility of the dripped neutrons.
This is opposite to our naive expectation.
In the density region where the slab phase is expected
($0.07\leq n_B<0.08$ fm$^{-3}$) at beta equilibrium,
the effective mass $\bar{m}_z^*$ is significantly smaller than
the bare mass, $m_z^*/m_n = 0.65\sim 0.75$.

Figure~\ref{fig:effective_mass} shows the effective masses at
the beta-equilibrium condition.
However, away from the beta-equilibrium condition,
these values can be much larger or smaller.
For instance, at $n_B=0.04$ fm$^{-3}$ and $Y_p=0.25$,
we obtain
$m_z^*/m_n=16.2$, $\tilde{m}_z^*/m_n = 2.8$, and
$\bar{m}_z^*/m_n = 0.53$.
The average effective masses are about 30 times different,
depending on their definition.
This is because only a small fraction of neutrons are dripped
in this case and the valence bands of $e^{(n)}_{\alpha,k_z}>U_n^0$
provide small effective masses.
As we see in Fig.~\ref{fig:effective_mass}, the effective masses
significantly vary from one band to another.
Thus, depending on which band the neutrons belong to,
they may differently respond to an external force.

\begin{figure}[tb]
\begin{centering}
\includegraphics[width=0.90\columnwidth]{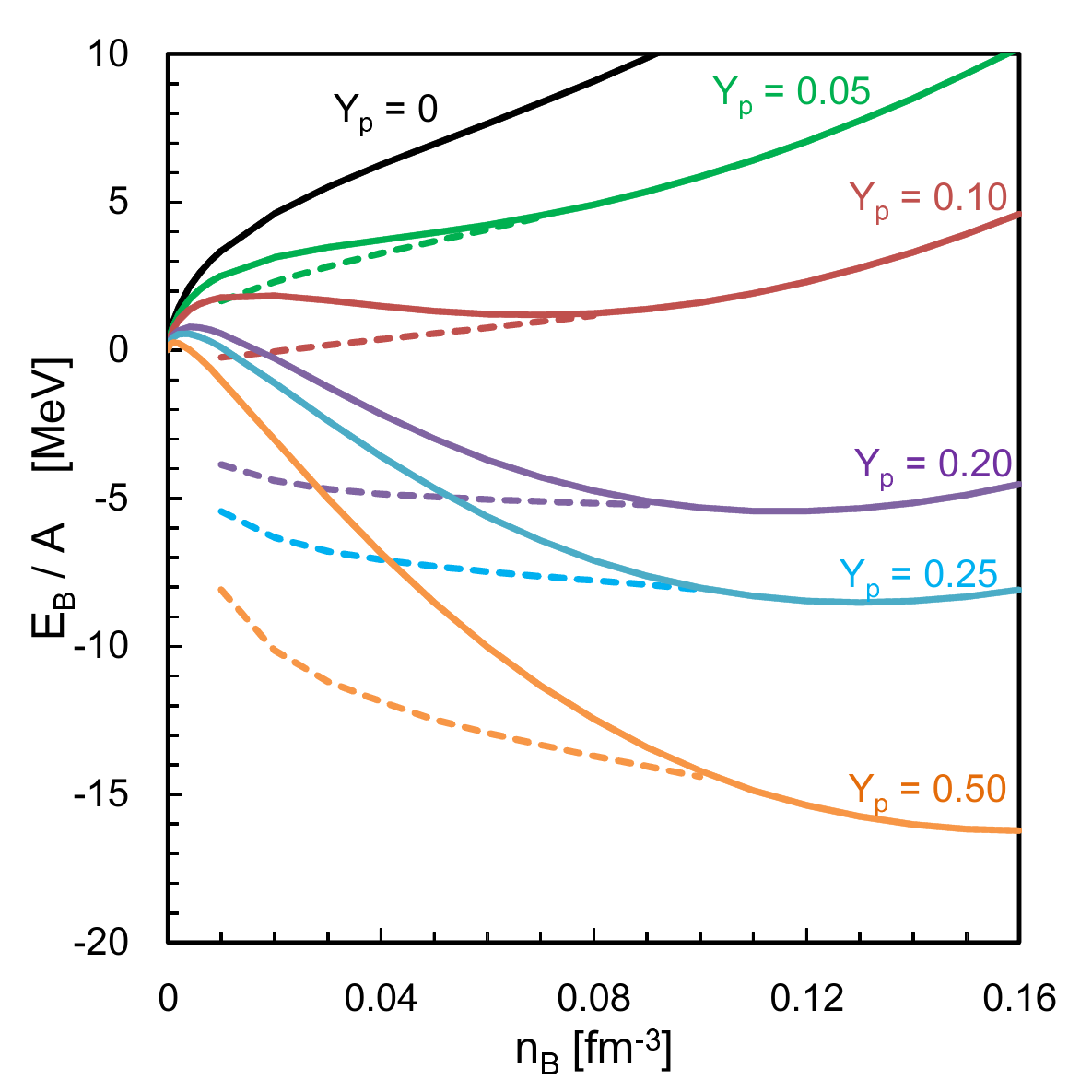}
\par\end{centering}
\caption{
(Color online)
Nuclear energy per baryon $E_B/A$ as functions of density $n_B$
for different values of $Y_p$.
The dashed lines indicate $E_B/A$ of the slab nuclei, while
the solid lines correspond to the uniform matter EOS.
}
\label{fig:EOS}
\end{figure}
\begin{figure}[tb]
\begin{centering}
\includegraphics[width=0.90\columnwidth]{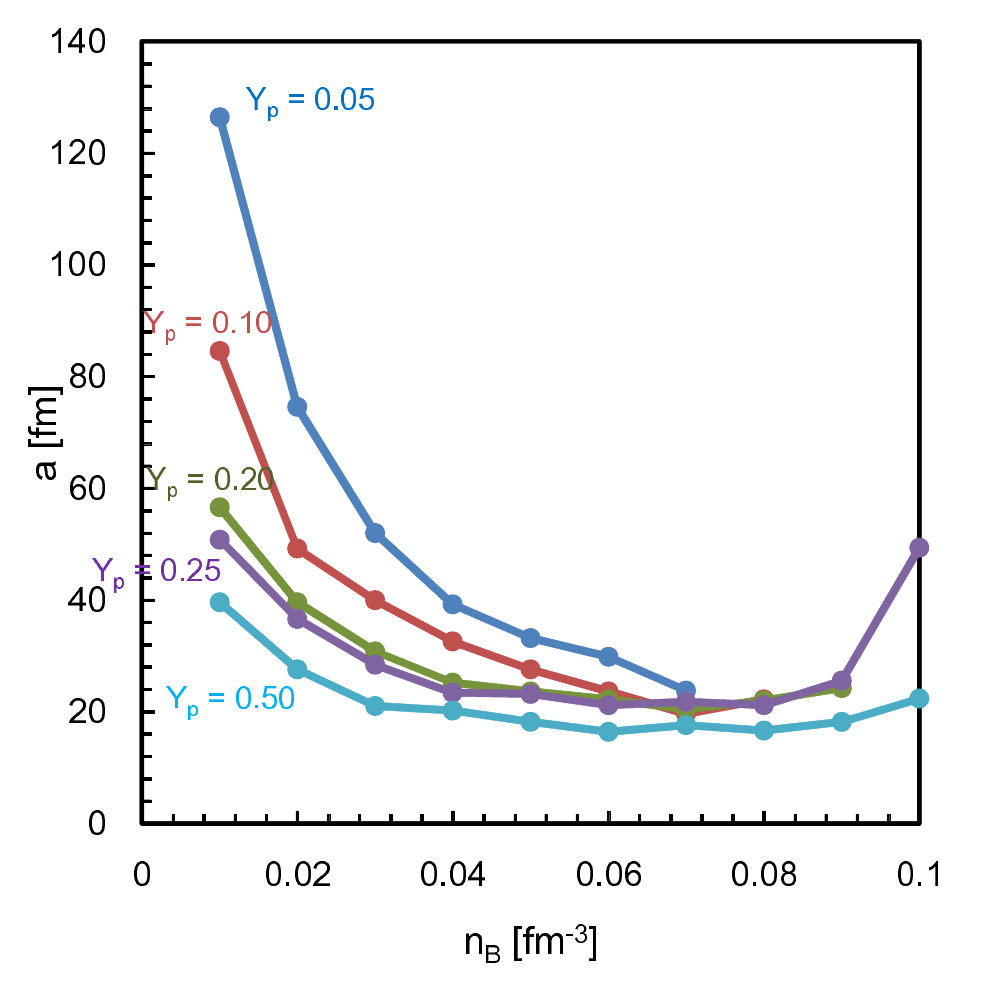}
\par\end{centering}
\caption{
(Color online)
Lattice constant $a$ as a function of density $n_B$.
Different lines correspond to different values of proton ratio $Y_p$.
}
\label{fig:lattice_const}
\end{figure}

\subsubsection{EOS of the slab phase}

Calculated energy per baryon is summarized in Fig.~\ref{fig:EOS}.
The non-uniform slab phase is favored at low density $n_B<0.1$ fm$^{-3}$.
Especially at large $Y_p$, the non-uniform structure is strongly favored
by reducing the Coulomb energy.
At the lowest density, difference in $E_B/A$ between these two phases
amounts to about 7 MeV.
Of course, in such low-density region,
we expect that other non-uniform phases,
such as rod or droplet, are even more favored.
The slab phase with $Y_p=0.5$ can be energetically most favored
among various phases of nuclear pasta,
in the density region of $0.05<n_B<0.07$ fm$^{-3}$ \cite{OMYT13}.
In this region, the energy gain is about 3 MeV per nucleon which
is still significant.
In contrast,
the beta-equilibrium states with $Y_p=0.026$
at $n_B=0.03$ fm$^{-3}$
gain energy of $\Delta E_B/A\approx 862$ keV.

The obtained lattice constant $a$ is shown in Fig.~\ref{fig:lattice_const}.
Around the density of $0.07\sim 0.08$ fm$^{-3}$,
the calculated lattice constant is $a=20-30$ fm,
irrespective of proton ratio $Y_p$.
The lattice constant increases in both low-density and high-density sides.
In the high-density side, in the vicinity of transition to uniform phase,
the anti-slab-like structure appears and $a$ increases.
In the low-density side, $a$ is significantly enhanced for
systems with small $Y_p$.
This can be understood as follows:
For symmetric or near symmetric case with $Y_p\sim 0.5$,
each slab has normal nuclear density and
the neutrons never drip from the slab.
Therefore, the lattice constant $a$ is basically determined by density $n_B$.
The Coulomb energy favors larger values of $a$, but the given density $n_B$
forbids $a$ from being too large.
In contrast, for small values of $Y_p$, since there exist dripped neutrons,
a larger value of $a$ is allowed at a given density $n_B$.

\section{Summary}
\label{sec:summary}

The fully self-consistent band calculation based on the BCPM
energy density functional has been performed for the slab phase
of the inner crust of neutron stars.
The lattice constant $a$ was determined by minimization of the total
energy for each value of density $n_B$ and proton ratio $Y_p$.
Comparing the result with that of Thomas-Fermi (TF) approximation,
we have found that the TF qualitatively reproduces the slab structure of
the self-consistent band calculation under
the beta-equilibrium condition.
This is partly because the slab nuclei have a shell effect weaker than
the other phases, such as rod and droplet.
However, away from the beta equilibrium, the lattice constant can
be significantly different, especially near the boundary between
the slab and uniform phases.
The Weizs{\"a}cker term in the TF theory plays an important role.
Without this term, the results even more deviate from the band calculation.

The band structure of neutrons are obtained from the present
calculation.
The calculated band gaps at $k_z=0$ and $\pm \pi/a$ are small,
typically, order of keV to tens of keV.
These values of the band gap are significantly smaller than
magnitude of neutron pairing gap, which is expected
to be order of hundreds of keV or MeV.
The non-dissipative entrainment effect is studied by calculating
the macroscopic effective mass due to the Bragg scattering.
We have found that, the calculated mobility coefficient $\mathcal{K}^{zz}$
and equivalently the conduction neutron density $n_z^c$,
are certainly reduced from the values for the uniform nuclear matter
($n_z^c < n_n$).
However, the average effective mass of dripped neutrons
$\bar{m}_z^*$ is smaller than the bare neutron mass.
It is somewhat surprising that,
the Bragg scattering enhances
the mobility of dripped neutrons.

In former studies \cite{CCH05,Cha12},
the calculated effective mass is always larger than the bare mass,
$m^*/m_n>1$, in all the (pasta) phases of the inner crust.
Their effective mass in the slab phase corresponds to $\tilde{m}_z^*$
of Eq. (\ref{effective_mass_2}), whose values are consistent with
our results, $\tilde{m}_z^*/m_n = 1.0-1.05$ in the density region
of $n_B \geq 0.07$ fm$^{-3}$.
In this definition, the ``free'' neutrons include those in
orbitals with $e^{(n)}_{\alpha, k_z}<U_n^0$
which are trapped inside the slab nuclei.
Therefore, we suppose that the mobility of neutrons dripped from the slabs
is better represented by the effective mass $\bar{m}_z^*$
rather than $\tilde{m}_z^*$.

We have performed the calculation for the one-dimensional slab phase
without the pairing correlations.
For the slab phase in the
inner crust of neutron stars, we can conclude that
the entrainment effect for neutrons
does not influence the conventional interpretation
of the pulsar glitches \cite{AI75,LEL99,AGHE12}.

This is the first step for the fully self-consistent band calculation
for the inner crust of neutron stars.
Obviously, its extension to rod-like and crystalline nuclei with
superfluid neutrons is highly desirable.
In addition, there is an recent argument that the effect of the band
structure is significantly hindered when the pairing gap $\Delta_n$ is
greater than the band gap $\Delta\epsilon_n$ \cite{WP17}.
The present calculation for the slab phase is exactly the case,
namely $\Delta\epsilon_n \ll \Delta_n$.
In order to take account of the pairing and the temperature effect
simultaneously,
we are currently developing a parallelized computer program of
the finite-temperature Hartree-Fock-Bogoliubov calculation
with the three-dimensional coordinate-space representation.
We plan to use this for our future studies of various phases
in the inner crust.


\begin{acknowledgments}
We thank Dr Kei Iida for valuable discussion in the initial stage
of the present work.
This work is supported in part by JSPS KAKENHI Grant No.18H01209,
and also by JSPS-NSFC Bilateral Program for Joint Research Project on
Nuclear mass and life for unravelling mysteries of r-process.
This research in part used computational resources provided by
Multidisciplinary Cooperative Research Program
in Center for Computational Sciences, University of Tsukuba.
\end{acknowledgments}

\bibliography{nuclear_physics,current,myself}

\end{document}